\documentclass[epj]{svjour}
\usepackage[dvipsnames]{xcolor}
\usepackage[breaklinks,colorlinks,urlcolor=blue,citecolor=blue]{hyperref}
\usepackage{graphicx}
\usepackage{amsmath}
\usepackage[normalem]{ulem}

\def\simle{\mathrel{\rlap{\raise 0.511ex
        \hbox{$<$}}{\lower 0.511ex \hbox{$\sim$}}}}
\def\simge{\mathrel{\rlap{\raise 0.511ex
       \hbox{$>$}}{\lower 0.511ex \hbox{$\sim$}}}}

\newcommand*{\Mmax}{\ensuremath{M_{\max}}}
\newcommand*{\Msun}{\ensuremath{\mathrm{M}_\odot}}

\begin{document}

\title{Neutron Star Radii, Universal Relations, and the Role of Prior
  Distributions}
\subtitle{}
\author{A.~W. Steiner\inst{1,2}, J.~M. Lattimer\inst{3},
  E.~F. Brown\inst{4,5,6}}
\authorrunning{Steiner, Lattimer, \& Brown}
\titlerunning{Neutron Star Radii}
\institute{
  Department of Physics and Astronomy, University of
  Tennessee, Knoxville, TN 37996, USA 
  \and
  Physics Division, Oak Ridge National Laboratory, Oak
  Ridge, TN 37831, USA
  \and
  Dept. of Physics \& Astronomy,
  Stony Brook University, Stony Brook, NY 11794-3800, USA
  \and
  Department of Physics and Astronomy, Michigan State University, East
  Lansing, MI 48824, USA
  \and
  The Joint Institute for Nuclear Astrophysics-Center for the
  Evolution of the Elements, Michigan State University, East Lansing,
  MI 48824, USA
  \and
  National Superconducting Cyclotron Laboratory, Michigan State
  University, East Lansing, MI 48824, USA
}
\date{Received: date / Revised version: date}

\abstract{We investigate constraints on neutron star structure arising
  from the assumptions that neutron stars have crusts, that recent
  calculations of pure neutron matter limit the equation of state of
  neutron star matter near the nuclear saturation density, that the
  high-density equation of state is limited by causality and the
  largest high-accuracy neutron star mass measurement, and that
  general relativity is the correct theory of gravity. We explore the
  role of prior assumptions by considering two classes of equation of
  state models. In a first, the intermediate- and high-density
  behavior of the equation of state is parameterized by piecewise
  polytropes. In the second class, the high-density behavior of the
  equation of state is parameterized by piecewise continuous line
  segments. The smallest density at which high-density matter appears
  is varied in order to allow for strong phase transitions above the
  nuclear saturation density. We critically examine correlations among
  the pressure of matter, radii, maximum masses, the binding energy,
  the moment of inertia, and the tidal deformability, paying special
  attention to the sensitivity of these correlations to prior
  assumptions about the equation of state. It is possible to constrain
  the radii of $1.4~\Msun$ neutron stars to a be larger than 10 km,
  even without consideration of additional astrophysical observations,
  for example, those from photospheric radius expansion bursts or
  quiescent low-mass X-ray binaries. We are able to improve the
  accuracy of known correlations between the moment of inertia and
  compactness as well as the binding energy and compactness. We also
  demonstrate the existence of a correlation between the neutron star
  binding energy and the moment of inertia.}


\maketitle

\section{A Brief History of Neutron Star Radii}
\label{sec:short-history}

Theoretical estimates of radii of spherically-symmetric non-rotating
neutron stars came directly from analytic and numerical solutions of
the Tolman-Oppenheimer-Volkov
equations~\cite{Tolman39ss,Oppenheimer39om}. At this time, the
critical role played by the maximum mass was already known. Twenty
years later, improved estimates of radii came directly from
improvements in nuclear physics: Ref.~\cite{Cameron59ns} used a
recently developed zero-range nucleon-nucleon interaction (the Skyrme
interaction~\cite{Skyrme59te}) to compute neutron stars with masses up
to a maximum of $2~\Msun$ and radii as small as 8 km and even larger
than 27 km. It was not realized at this time that the EOS was acausal
at high densities, thus leading to extremely small radii. Also, the
extremely large radii were found for stars with masses below what is
commonly considered to be the minimum formation mass of neutron stars,
about $1~\Msun$. Later calculations of the EOS of neutron star matter
using both Skyrme and relativistic field-theoretical interactions
constrained by nuclear experiment have found radii between about 9 and
15 km~\cite{Steiner05ia}. Astronomical observations of radio
pulsations in neutron stars followed 8 years later in 1967;
observations of pulsars in binary systems have lead to numerous mass
measurements (e.g., see Ref.~\cite{Phinney94}). Importantly, the
largest accurately measured masses~\cite{Demorest10,Antoniadis13} set
a lower limit to the neutron star maximum mass, which, combined with
general relativity and causality, constrain minimum values for neutron
star radii~\cite{Lattimer12} of all masses. For a $1.4~\Msun$ star,
the current minimum maximum mass $M_{\mathrm{max}}=2~\Msun$
establishes that its radius $R_{1.4}>8.15$ km. Larger mass
measurements will increase this minimum radius. An important
theoretical development was the realization that neutron star radii
for stars in the mass range $1~\Msun<M<1.8~\Msun$ are primarily
determined by the EOS in the density range
$n_s<n<2n_s$~\cite{Lattimer01}, where $n_s$ is the nuclear saturation
density. This provides a direct link between both nuclear experiment
and theory and neutron star radii (though the link is weaker if there
is a strong phase transition in this density range, as demonstrated
below). One of the objectives of this paper is to use recent
developments in nuclear experiment and theory, coupled with the
assumption that neutron stars have crusts, to establish neutron star
radius constraints. In addition, we investigate how and whether
astrophysical observations of neutron stars in X-rays can
realistically further improve these limits.

Ref.~\cite{vanParadijs79} proposed using photospheric radius expansion
(PRE) X-ray bursts to obtain simultaneous mass and radius measurements
in 1979, but the method did not lead to interesting constraints until
2006~\cite{Ozel06}. Masses and radii from PRE X-ray
bursts~\cite{Guver10a,Guver10b,Ozel09b} gave radii between 8 and 12 km
and masses between 1.2 and 2 $\Msun$ under the assumption that the
photosphere at ``touchdown'' has receded to the neutron star surface.
Under the assumption that this model describing PRE X-ray bursts was
correct, large radii and EOS models with high pressures between $n_s$
and $2n_s$ seemed to be ruled out.

Observations of thermal emission from accreting neutron stars in
binary systems, known as quiescent low-mass X-ray binaries (qLMXBs),
also provide radius measurements. The earliest such observations led
to radius estimates of less than 1 km, much smaller than that
predicted from theoretical models~\cite{vanParadijs87sx}. For
accreting neutron stars, the photosphere is expected to consist of
hydrogen \cite{bildsten92}, and Ref.~\cite{Rutledge99tt} showed that
models of thermal emission from pure hydrogen photospheres
\cite{rajagopal96,zavlin96} gave inferred radii on the order of 10 km.
Neutron star radius information from quiescent low-mass X-ray binaries
began to provide significant constraints on radii in 2006 (at the same
time as PRE X-ray bursts), due to improved observations and a
consistent treatment of the surface gravity in the neutron star
atmosphere model used to fit the spectrum~\cite{Heinke06,Webb07}. This
led to refined estimates of radii between 8 and 16 km, which, however,
by themselves did not rule out a significant number of contemporary
theoretical EOS models~\cite{Lattimer01}.

There were two principal difficulties which prevented these
measurements from tightly constraining the EOS. The first difficulty
was the fact that taking neutron star mass and radius measurements and
constructing an EOS formally requires an inversion of the TOV
equations. A method to invert the TOV equations was developed in
1992~\cite{Lindblom92}, but this approach is subject to numerical
difficulties (see recent work in Ref.~\cite{Lindblom14}). The second
difficulty was that the traditional approach of performing a
chi-squared fit is prohibited by the large uncertainties and the
potential for the mass-radius curve to be non-monotonic in both mass
and radius (see a more recent summary of the possible mass-radius
curve shapes in Ref.~\cite{Alford13}). Ref.~\cite{Ozel09} showed that,
if one chooses a sufficiently generic EOS parameterization, a formal
inversion is unnecessary. Ref.~\cite{Read09} showed that a
three-parameter model based on piecewise polytropes (commonly used,
e.g., in numerical solutions of rotating relativistic
stars~\cite{Eriguchi85}) is general enough to reproduce modern
theoretical EOSs to within a few percent. Thus Ref.~\cite{Ozel09} used
this parameterization to show that, given simultaneous mass and radius
measurements of three neutron stars, one can marginalize over the
unknown neutron star masses to obtain probability distributions for
the three unknown EOS parameters. This was applied in
Ref.~\cite{Ozel10} to three simultaneous mass and radius constraints
for three neutron stars. For the EOS near and below the nuclear
saturation density (assumed to have zero uncertainty),
Ref.~\cite{Ozel10} used the Skyrme interaction SLy4 that was fit to
nuclei and theoretical calculations of low-density neutron
matter~\cite{Chabanat95ns}. It was claimed that the mass and radius
measurements presented a challenge for theories of neutron star
matter, but it is now understood that systematic uncertainties in the
X-ray burst model are the more likely culprit for the tension
that was observed.

Soon after, Ref.~\cite{Steiner10te} used Bayesian inference to combine
mass and radius constraints from both qLMXBs and PRE X-ray bursts to
obtain constraints on the EOS. This work used an alternate nuclear
physics-based parameterization for matter near the nuclear saturation
density and piecewise polytropes at higher densities (in a slightly
different form than that presented in Ref.~\cite{Read09}). In the case
that the prior distributions for the polytrope parameters are uniform,
the Bayesian inference-based method to obtain the EOS parameters is
similar to the method developed in Ref.~\cite{Ozel09}, except that
Ref.~\cite{Steiner10te} additionally obtained probability
distributions for the mass-radius curve by marginalizing over the
posteriors for the radius as a function of mass. Also, the direct use
of Bayesian inference in Ref.~\cite{Steiner10te} allowed the use of
more parameters which more fully explored the uncertainties in the EOS
near saturation densities and led to constraints on the density
dependence of the nuclear symmetry energy. Finally,
Ref.~\cite{Steiner10te} showed that the assumption that the
photosphere at touchdown is coincident with the surface (i.e., not
extended) is not consistent with the data (within the context of the
model being used for PRE X-ray bursts in Refs.~\cite{Steiner10te} and
\cite{Ozel10}). The final result was a radius range of 10.7 to 12.5 km
for a 1.4 $\Msun$ neutron star. Radii smaller than 10.7 km were ruled
out, in contrast to earlier results based on PRE X-ray bursts. These
improved radius constraints came principally as a result of the
marginalization from Bayesian inference. However, it was reported in
Ref.~\cite{Steiner10te} that there were several systematic
uncertainties which potentially invalidated this result.

One important systematic uncertainty was the potential for accretion
to muddle the interpretation of PRE X-ray bursts. In 2011,
Ref.~\cite{Suleimanov11} used longer bursts and a more complete model
of the neutron star atmosphere to obtain a radius greater than 14 km
for the single source studied. For the qLMXB sources, the composition
of the atmosphere plays an important role. Ref.~\cite{Servillat12}
showed that a helium (rather than hydrogen) atmosphere changes the
inferred radius range for the neutron star in M28 from 6 to 11.5 km to
7 to 17 km. Both of these systematic uncertainties continue to play a
role in interpreting mass and radius measurements. At around the same
time, work progressed on constraining the nature of the
nucleon-nucleon interaction from neutron star mass and radius
observations~\cite{Hebeler10b,Steiner12cn}, showing that current
observations suggested a relatively weak density dependence in the
nuclear symmetry energy, predicting radii less than 14 km.

In 2013, Ref.~\cite{Steiner13tn} re-analyzed the qLMXB and PRE data
including some systematic uncertainties ignored in previous works,
including variations in the color correction factor used to infer
masses and radii from the PRE X-ray bursts. This systematic
uncertainty extended the radius range from 10.4 to 12.9 km. Another
critical systematic included in Ref.~\cite{Steiner13tn} is the
ambiguity in the choice of the prior distribution. Marginalizations
over model parameters include a choice of prior distribution, whether
implicit as in Refs.~\cite{Ozel09,Ozel10} or explicit as in
Ref.~\cite{Steiner10te}. The ambiguity from the choice of prior is not
important if the number of parameters is sufficiently small. Also,
prior assumptions are not important if uncertainty in the parameter
that represents the independent variable of the model function (e.g.,
the mass of the neutron star) is small. If both of these conditions
held, one could perform the classical chi-squared fit and directly
obtain the mass-radius curve and the associated EOS. However, these
conditions did not then (and still do not) hold for neutron star mass
and radius observations.

One way the choice of prior distribution is important is the extent to
which it favors or disfavors the presence of phase transitions where
the pressure is nearly flat with increasing density. Polytropic EOSs,
when used with uniform prior distributions for the polytropic
exponent, naturally disfavor phase transitions. The reason for this is
that all polytropes go through the origin, thus the polytropic
exponent must be very small in order to produce a nearly flat pressure
as a function of energy density (which would result from a strong
phase transition). Ref.~\cite{Steiner13tn} found that this prior
ambiguity was also important and also that the 14 km radius obtained
in Ref.~\cite{Suleimanov11} was difficult to reconcile with the
smaller radii obtained from qLMXB sources.

Ref.~\cite{Guillot13} combined new qLMXB observations and a constant
radius model (motivated by the vertical shape of the mass-radius
curves obtained in Ref.~\cite{Steiner13tn}) to obtain a very small
preferred neutron star radius with rather small error, $9.0\pm1.4$  km.
However, when individually analyzed, the sources produced predicted
radii in a wide range (7 to 20 km). Ref.~\cite{Lattimer14co} concluded
that two systematic uncertainties were responsible for this result:
(i) some of the small radius neutron stars may have helium rather than
hydrogen atmospheres (as discussed above) and (ii) uncertainties in
the hydrogen column density may give incorrect radii.
Ref.~\cite{Heinke14} confirmed that the hydrogen column density
inferred for one neutron star in the data set was indeed
systematically shifted. Ref.~\cite{Heinke14} also showed that the
choice of galactic abundance model was an important systematic.

Ref.~\cite{Lattimer14co} also introduced the use of the Bayes factor
\footnote{The Bayes factor is sometimes referred to as the ``odds
  factor'' or ``likelihood ratio''. The latter term, however, is often
  used to describe the ratio of the maximum likelihoods, rather than
  the ratio of the integrals.} to decide between various models. In
the case of a chi-squared fit, the model with the smaller value of
chi-squared gives a better fit. In the frequentist approach, an
e\-qui\-va\-lent method is the max\-i\-mum like\-li\-hood test: the
pre\-fer\-red mo\-del is the one with the largest maximum likelihood.
A test using the Bayes factor is the Bayesian analog of the maximum
likelihood test. The Bayes factor for model A versus model B is a
ratio of the ``evidence''. The evidence, in turn, is defined as the
integral over the posterior (see Ref.~\cite{Lattimer14co} for a brief
review and Ref.~\cite{Steiner15mb} for additional discussion).

Bayesian inference continues to be an important tool for analyzing
neutron star mass and radius observations, in part because the
uncertainties are still large and the problem of determining the
mass-radius curve (or the EOS) is underconstrained. For this reason,
the Bayes factor remains one of the best tools to compare the evidence
for (or against) the various models and assumptions which are
employed. Models with tightly constrained posteriors are not preferred
because they correspond to small values of the evidence. One way to
think about this result is to note that models with tightly
constrained posteriors are too finely-tuned because there are only a
very few subset of model parameters for which the model is not ruled
out.

More recently, Ref.~\cite{Lattimer14co} showed that Bayes factors
imply that an extended photosphere at touchdown is preferred in PRE
X-ray bursts (the posteriors for an extended photosphere are broad and
thus that the evidence for the model is large) and helium, rather than
hydrogen, atmospheres are favored in some qLMXBs. However, this work
did not rule out the potential for other interpretations and
Ref.~\cite{Ozel15a} has since suggested that rotation and temperature
corrections to the Eddington limit may also play a role.

Nevertheless, Ref.~\cite{Guillot14}, implementing new data which
resolved some difficulties in the previous study~\cite{Guillot13}, but
retaining their assumptions of a common radius and hydrogen
atmospheres for all sources, continued to obtain a small radius
($9.4\pm1.2$ km).

\section{The Role of Prior Distributions}
\label{s:2}

The choice of prior distribution continues to play an important role
in interpreting observations and theory. This statement trivially
holds true, as in the context of Bayesian inference all model
assumptions can be viewed as a particular choice of prior
distribution. Nevertheless, it is important when one can quantify the
effect that these assumptions have. A recent analysis of both the
qLMXB and PRE data in Ref.~\cite{Steiner15un} gives two different
ranges for the radius of a 1.4 solar mass neutron star under different
prior assumptions. Model A (based on polytropes) gave a radius range
of 10.8 to 12.4 km while Model C (which allows for stronger phase
transitions) gave a radius range of 10.2 to 11.9 km (this distinction
was first observed in Ref.~\cite{Steiner13tn}). Since the Bayes factor
for Model A versus Model C is not significantly different from 1, one
must presume that either model could be correct, and the full range of
allowed radii for a 1.4 $\mathrm{M}_{\odot}$ neutron star is between
10.2 and 12.4 km. The prior ambiguity increases the uncertainty by at
least 30\%. The constraints on the pressure at twice the saturation
density, as inferred from neutron star radius measurements, are also
sensitive to the prior distribution: Model A gives a range between 9.1
and 23.0 $\mathrm{MeV}/\mathrm{fm}^3$ while Model C gives a range
between 2.3 and 17.0 $\mathrm{MeV}/\mathrm{fm}^3$~\cite{Steiner15un}.
Had one considered only a polytrope-based parameterization of the EOS
one would have overestimated the lower limit on the pressure by a
factor of four. Recently, Ref.~\cite{Nattila15} found a similar
  variation in the radius when analyzing X-ray bursts in the hard
  state. The final result was 11.3 to 12.8 km for a 1.4
  $\mathrm{M}_{\odot}$ neutron star for Model A and 10.5 to 12.5 km
  for Model C.

Ref.~\cite{Ozel15b} compared two different methods for obtaining
masses and radii from PRE X-ray bursts. The first method is basically
the method given in Ref.~\cite{Ozel10} and Ref.~\cite{Steiner10te}. In
this first method, the marginalization over model parameters includes
a Jacobian factor which transforms between a two-dimensional space
over touchdown flux and normalization to a two-dimensional space over
mass and radius. In the second method of Ref.~\cite{Ozel15b}, this
Jacobian is removed. Ref.~\cite{Ozel15b} confusingly refers to their
first method as the ``frequentist'' approach and the second method as
a ``Bayesian'' approach, even though the presence of the Jacobian is
simply a transformation of variables (and is widely used by Bayesian
practitioners~\cite{Gelman14}). In this article, we employ a language
that is more typical in the statistics literature. In
Ref.~\cite{Ozel15b}, these two methods are compared by a qualitative
examination of ``bias'' in the method. Bias is measured by comparing
the posterior in the mass-radius space distribution to the prior
distribution in mass-radius space. 

A quantitative approach which is more similar to the language of the
statistics literature would compute the evidence for these two methods
and form the ratio to compute the Bayes factor. The Bayes factor is
not the same as the ``bias'' used in Ref.~\cite{Ozel15b}: it measures
the integral under the posterior rather than its location in parameter
space relative to the prior. In the case that the relative Bayes
factor between these two models is near unity both methods would need
to be considered, just as was done with Models A and C in
Ref.~\cite{Steiner15un} and also below. 

Prior distributions are also relevant for determining the nature of
``correlations'' or ``universal relations''. There have been several
correlations observed within nuclear structure and be\-tween
nu\-cle\-ar struc\-ture and neutron stars: the correlation between the
EOS of neutron matter and the neutron skin thickness of
lead~\cite{Typel01}, the correlation between the pressure of
neutron-rich matter and neutron star radii~\cite{Lattimer01}, and the
correlation between the neutron skin thickness of lead and neutron
star radii~\cite{Horowitz01}. These correlations are principally
explored by studying a range of several model parameterizations. They
reflect the nature of the uncertainties in two quantities vis \`{a}
vis the current knowledge of the nucleon-nucleon interaction. In the
context of Bayesian inference, these correlations describe the nature
of the posterior distribution of two quantities given a particular
prior distribution. The shape of the correlation is thus dependent on
the prior distribution\footnote{Alternatively, in the context of a
  hierarchical Bayesian analysis, the shape of the correlation is
  dependent on the hyper-prior distribution of hyper-parameters.}.

In the past several years, other kind of correlations within neutron
stars, which have been referred to as ``universal relations'', e.g.,
the correlation between the moment of inertia and the Love number of a
neutron star~\cite{Yagi13}. The uncertainties in these universal
relations are also dependent on the prior assumptions, though the
magnitude of this prior dependence can be very different from one
correlation (or one universal relation) to another.

\section{The Equation of State}
\label{s:3}

\subsection{Neutron Star Crusts and the Low-Density EOS}
We assume that neutron stars have crusts, {\it i.e.}, they have a
surface region with densities less than approximately $10^{14}$ g
cm$^{-3}$ composed largely of nuclei, neutrons and electrons in beta
equilibrium~\cite{Baym71}. The pressure in this region is largely due
to relativistic, degenerate electrons with at most a 5\% contribution
from nuclei and neutrons. Since the nuclei are in pressure equilibrium
with the neutrons, they individually contribute almost no pressure
since their baryon density is close to the nuclear saturation density
$n_s\simeq0.16$ fm$^{-3}$ or $\rho_s\simeq2.7\times10^{14}$ g
cm$^{-3}$ where uniformly dense symmetric matter has zero pressure.
The major contribution of baryons to the pressure is from the
collective Coulomb pressure due to the nuclear lattice, and is
therefore largely independent of uncertainties in the equation of
state of nuclear matter. Various calculations indicate that the
transition from crustal material to uniform nuclear matter, $n_t$,
occurs in the range $n_s/4-n_s/2$.

Recent calculations of the properties of pure neutron matter have
produced estimates of the pressure-energy density relation up to about
$2n_s$. Matter in neutron stars at densities between $n_t$ and $2n_s$
is extremely neutron-rich because it is in beta equilibrium.
This condition is equivalent to minimization of the total energy per
baryon with respect to the charge fraction $x=n_p/n$ where $n_n$ and
$n_p$ are the neutron and proton baryon densities, respectively, and
$n=n_n+n_p$. The difference between the energy of pure neutron matter
and symmetric matter (with equal numbers of neutrons and protons) is
called the nuclear symmetry energy $S(n)$, and the energies of
intermediate proton fractions can be approximated with a quadratic
interpolation between these extremes:
\begin{equation}
E(n,x)\simeq E_{1/2}(n)+S(n)(1-2x)^2,
\label{eq:e}
\end{equation}
where $E_{1/2}(n)$ is the energy per baryon of symmetric matter. A
crude estimate for the symmetry energy near $n_s$ is
\begin{equation}
S(n)=S_{v}(n/n_s)^\gamma; 
\label{eq:s}
\end{equation}
nuclear experimental information and neutron matter calculations
indicate that $26{\rm~MeV}\simle S_{v}\simle34{\rm~MeV}$ and
$0.3\simle\gamma\simle0.7$. The pressure corresponding to Eqs.
(\ref{eq:e}) and (\ref{eq:s}) is
\begin{eqnarray}
p(n,x)&=&n^2\frac{\partial E(n,x)}{\partial n}\simeq
p_{1/2}(n)+\nonumber \\
&& S_{v}\gamma n_s(n/n_s)^{\gamma+1}(1-2x)^2,
\label{eq:p}
\end{eqnarray}
where $p_{1/2}(n)$ is the pressure of symmetric matter. Note that, by
definition, $p_{1/2}(n_s)=0$; to leading order, near $n_s$, the
symmetric matter pressure increases linearly with density,
$p_{1/2}(n)\simeq(K_s/9)(n-n_s)$, where $K_s\simeq240$ MeV is the
nuclear incompressibility parameter.

As stated above, matter in neutron stars is in beta equilibrium:
\begin{equation}
\frac{\partial[E(n,x)+E_e(n,x)]}{\partial x}=\mu_p+\mu_e-\mu_n=0,
\label{eq:ex}
\end{equation}
where $E_e=(3/4)\hbar cx(3\pi^2nx)^{1/3}$ is the electron energy per
baryon assuming relativistic degenerate electrons, and the $\mu$'s are
chemical potentials. This is equivalent to
\begin{equation}
4S(n)(1-2x)=\hbar c(3\pi^2nx)^{1/3}.
\label{eq:be}
\end{equation}
This equation can be solved as a cubic equation for $x$ at a specific
density with the approximate solution
\begin{equation}
  x\simeq\frac{64S(n)^3}{3\pi^2(\hbar c)^3+128S(n)^3}
  \label{eq:be1}
  \end{equation}
which has the value $x\simeq0.04$ when $n=n_s$. The pressure of pure
neutron matter with ansatz Eq.~(\ref{eq:s}), at $n_s$, is $p=\gamma
n_sS_{v}.$ In beta equilibrium, the approximate neutron star pressure
at $n_s$ is
\begin{equation}
  p(n_s,x_\beta)\simeq\gamma n_sS_v\left[1-
    \left(\frac{4S_{v}}{\hbar c}\right)^3\frac{4\gamma-1}{3\pi^2\gamma}\right].
  \label{eq:be2}
\end{equation}
The correction term in Eq. (\ref{eq:be2}) is of order 1.4\%, and can
be ignored to good approximation. At higher densities, the proton
fraction and the correction term generally increase due to the
increasing symmetry energy. There is also a contribution from
$p_{1/2}(n)$. However, for densities up to $2n_s$ the neutron star
matter pressure is essentially equivalent to pure neutron matter
pressure.

Therefore, for densities between the core-crust transition density and
about $2n_s$, experimental information about the symmetry energy and
calculations of pure neutron matter pressures offer viable constraints
on the equation of state. Experimental information concerning the
symmetry energy is usually encoded in the parameters $S_v$ and $L$,
defined as
\begin{align}
  S_v&\equiv\frac{1}{8}\left[\frac{\partial^2E(n,x)
    }{\partial x^2}\right]_{n_s,1/2}\\
  L&\equiv\frac{3}{8}\left[\frac{\partial^3E(n,x)
    }{\partial n\partial x^2}\right]_{n_s,1/2}.
  \label{eq:sl}
\end{align}
For the symmetry energy of Eq. (\ref{eq:s}), one finds $L=3\gamma S_v$
so that $0.9S_v\simle L\simle2.1S_v$.

\begin{table*}[t]
\begin{centering}
\begin{tabular}{c||cccc||ccccccccc}
\hline
Model & $a$ & $\alpha$ & $b$ & $\beta$ & $S_v$ & $L$ &  $p_1$ & $\gamma_1$ \\ 
& MeV && MeV && MeV & MeV  & MeV fm$^{-3}$& \\
\hline
GCR 0 & 12.7 & 0.49 & 1.78 & 2.26 & 30.5 & 31.3 & 7.272 & 2.113 \\
GCR 1 & 12.7 & 0.48 & 3.45 & 2.12 & 32.1 & 30.8 & 10.402 & 2.335 \\
GCR 2 & 12.8 & 0.488 & 3.19 & 2.20 & 32.0 & 40.6 & 10.537 & 2.343 \\
GCR 3 & 13.0 & 0.475 & 3.21 & 2.47 & 32.0 & 44.0 & 13.274 & 2.487 \\
GCR 4 & 12.6 & 0.475 & 5.16 & 2.12 & 33.7 & 51.5 & 14.304 & 2.533 \\
GCR 5 & 13.0 & 0.50 & 4.71 & 2.49 & 33.8 & 56.2 & 18.678 & 2.700 \\
GCR 6 & 13.4 & 0.514 & 5.62 & 2.436 & 35.1 & 63.6 & 20.933 & 2.770 \\
\hline
DSS 0 & 10.94 & 0.459 & 4.106 & 1.977 & 31.1 & 39.4 & 8.125 & 2.182 \\
DSS 1 & 11.00 & 0.460 & 4.425 & 1.947 & 31.4 & 41.0 & 8.453 & 2.206 \\
DSS 2 & 11.95 & 0.495 & 3.493 & 2.632 & 31.4 & 45.3 & 13.760 & 2.509 \\
DSS 3 & 11.02 & 0.460 & 4.683 & 1.935 & 31.7 & 42.4 & 8.768 & 2.229 \\
DSS 4 & 10.95 & 0.454 & 5.158 & 1.972 & 32.1 & 45.4 & 9.676 & 2.290 \\
DSS 5 & 10.34 & 0.429 & 4.954 & 2.024 & 31.3 & 43.4 & 9.180 & 2.258 \\
DSS 6 & 10.29 & 0.433 & 7.227 & 1.842 & 33.5 & 53.3 & 11.241 & 2.384 \\
\hline
\end{tabular}\label{tab:nm}
\caption{Neutron matter calculations fit to the energy
  parameterization of Eq. (\ref{eq:nm}). GCR are models from Gandolfi,
  Carlson \& Reddy~\cite{Gandolfi12}; DSS are models from Drischler,
  Soma \& Schwenk~\cite{Drischler14}.}
\end{centering}
\end{table*}

\subsection{Piecewise Polytropes For the High-Density EOS}
\label{s:piece}

In our first model, we assume the crust-core tran\-si\-tion
den\-si\-ty is $n_0=n_s/2.7$, and use the crust EOS from
Ref.~\cite{Baym71}, for which the pressure is $p_0=0.243$ MeV
fm$^{-3}$, the energy density is $\epsilon_0=56.39$ MeV fm$^{-3}$, and
the internal energy per baryon is $E_0=12.13$ MeV. For densities above
the core-crust transition density, a scheme similar to a three
piecewise polytrope scheme explored by Ref.~\cite{Read09} is used.
There are 7 parameters. Four parameters correspond to boundary density
points ($n_i,~i=0-4$), and three correspond to the polytropic
exponents ($\gamma_i$) of the region between the densities $n_i$ and
$n_{i-1}$. Read et al.~\cite{Read09} discovered that a variety of
equations of state were successfully approximated by piecewise
polytropes with a common set of boundary densities $n_1\simeq1.85n_s$
and $n_2\simeq3.7n_s$. Assuming that the core-crust transition
pressure $p_0$ and energy density $\epsilon_0$ are fixed by the
crustal equation of state, and the core-crust transition density $n_0$
is chosen to be $n_s/2.7$, fixes two more parameters. In this case,
the three remaining parameters are the polytropic exponents
$\gamma_i$. We choose, equivalently, the pressures $p_1, p_2$ and
$p_3$ at the boundaries as parameters, following Ref.~\cite{Ozel09}.
The polytropic exponents are given by
\begin{equation}
  \gamma_i=\frac{\ln(p_i/p_{i-1})}{ \ln(n_i/n_{i-1})},
  \label{eq:ge}
\end{equation}
for $i=1\textrm{--}3$. Assuming continuity of both energies and
pressures at the boundary points $n_i$, the density and energy density
within $n_{i-1}<n<n_i$, for $i=1-2$, and for $n_{i-1}<n$, for $i=3$,
are
\begin{eqnarray}
  n&=&n_{i-1}\left(\frac{p}{p_{i-1}}\right)^{1/\gamma_i},\\
  \epsilon&=&\frac{p}{\gamma_i-1}+\left(\epsilon_{i-1}-
          \frac{p_{i-1}}{\gamma_i-1}\right)\frac{n}{ n_{i-1}},
          \label{eq:ge2}
\end{eqnarray}
where $\epsilon_{i-1}=n_{i-1}(m_n+E_{i-1})$ and $E_{i-1}$ are the
energy density and energy per baryon at the point $n_{i-1}$,
respectively.

\begin{figure}
  \includegraphics[width=8.5cm]{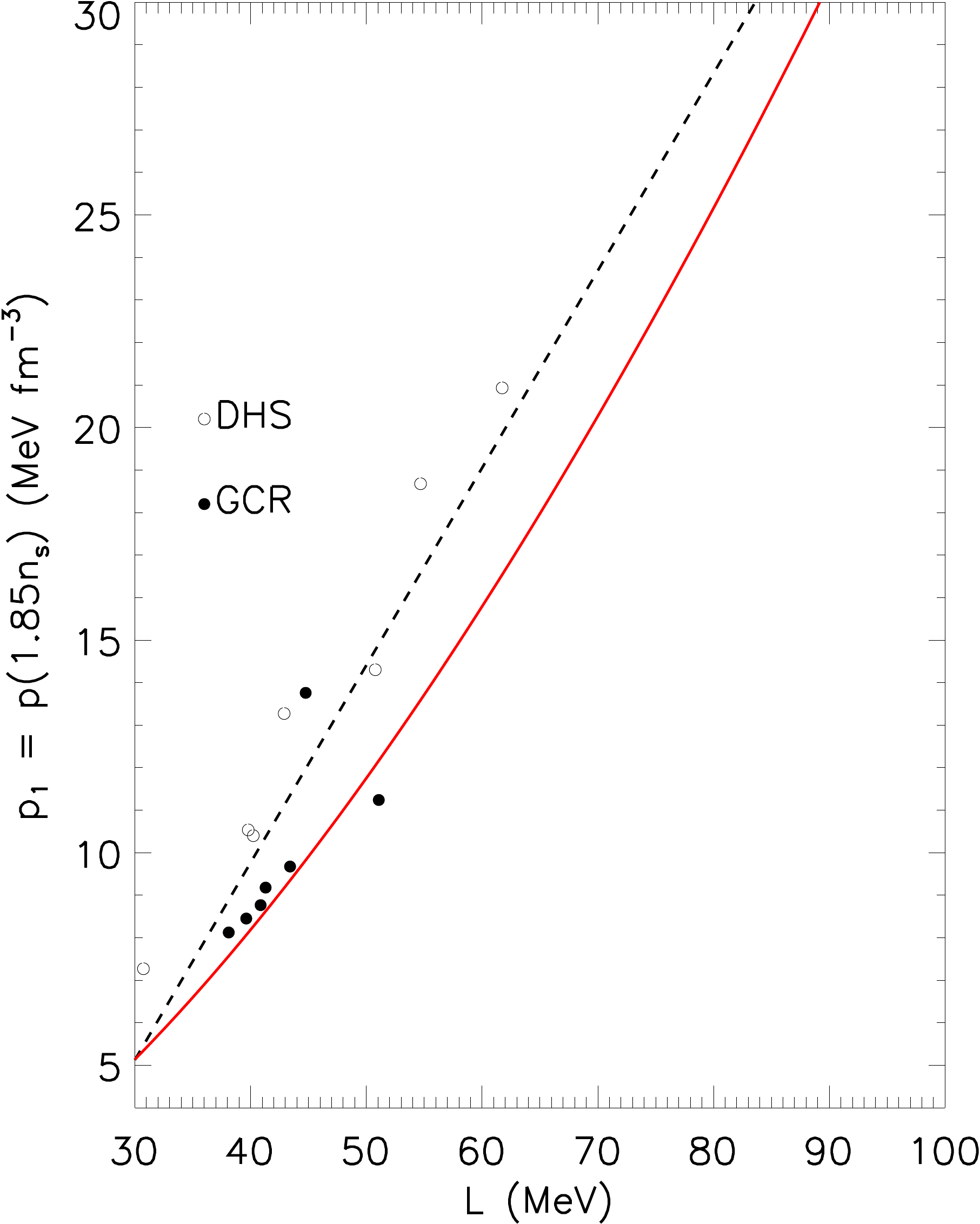}
  \caption{Values of $p_1=p(1.85n_s,0)$ and $L$ for the neutron matter
    EOSs in Table \ref{tab:nm}: solid and open circles represent the
    results from Refs.~\cite{Gandolfi12} and \cite{Drischler14},
    respectively. The dashed line shows the collective correlation
    between the two quantities for the neutron matter EOSs. The red
    solid curve indicate the values of $L_{pw}$ resulting from the
    piecewise polytrope, Eq. (\ref{eq:lpw}).}
  \label{fig:p1l}
\end{figure}

The boundary $n_1$ is sufficiently close to $n_s$ that neutron matter
calculations, which are claimed to be reliable to such densities,
offer a viable method of estimating $p_1$ and $\epsilon_1$, and
therefore $\gamma_1$. We shall adopt the approach that $p_1$ is
bracketed by the range of neutron matter calculations performed by
Refs.~\cite{Gandolfi12} and \cite{Drischler14}. As noted in
Ref.~\cite{Lattimer01}, the neutron star matter pressure between $n_s$
and $2n_s$ essentially determines the radii of neutron stars in the
mass range $1~\Msun$ to $1.6~\Msun$ (as long as one assumes no strong
phase transitions in this region, see e.g. Fig.~\ref{fig:R14P1both}).
We therefore expect that the value of $p_1$ will play the same role in
this EOS. Refs.~\cite{Gandolfi12} showed that the neutron
matter energy for densities less than about $2n_s$ was adequately
approximated by the double power law
\begin{equation}
E(n,0)\simeq a(n/n_s)^\alpha+b(n/n_s)^\beta
\label{eq:nm}
\end{equation}
where $a, b, \alpha$ and $\beta$ are parameters. For this energy
formula, and assuming the validity of a quadratic symmetry energy
interpolation (Eq. \ref{eq:e}), we immediately find that $S_v=B+a+b$
and $L=3(a\alpha+b\beta)$, where $B\equiv-E(n_s,1/2)\simeq16$ MeV is
the bulk binding energy of symmetric matter. Table \ref{tab:nm}
displays parameter values found by Ref.~\cite{Gandolfi12} for quantum
Monte Carlo neutron matter calculations (see also
Ref.~\cite{Steiner12cn}). We have also displayed parameter values that
fit the neutron matter results of Ref.~\cite{Drischler14} up to
densities $\simeq1.5n_s$, and we will assume these calculations can be
extrapolated to the slightly higher density $n_1=1.85n_s$. For each
set, the corresponding values of $S_v$, $L$, and the pressure $p_1$ at
$n_1$ have been tabulated. In the quadratic approximation for the
isospin dependence of the nucleon energies, $L=3p(n_s,0)/n_s$, so
$p_1$ and $L$ are highly correlated (Fig. \ref{fig:p1l}). The
piecewise polytrope approximation for the EOS, assuming a value for
$p_1$, explicitly predicts a value for $L$,
\begin{equation}
  L_{pw} = \frac{3p_0}{ n_s}2.7^{\gamma_1}=\frac{3p_1}{ n_s}1.85^{-\gamma_1}
\label{eq:lpw}
\end{equation}
with $\gamma_{1}$ given by Eq.~(\ref{eq:ge}). These values are close
to, but generally larger than, the corresponding neutron matter
predictions for a given value of $p_1$, indicating that the piecewise
polytropic EOS is a reasonable approximate in this density range.

Since the parameter set GCR 0 corresponds to an EOS with no three-body
interactions, the range
\begin{align}
  p_{n,\min}&=8.125{\rm~MeV~fm}^{-3}< p_1\nonumber \\
  p_1&<20.933{\rm~MeV~fm}^{-3}=p_{n,\max}
\end{align}
is predicted from realistic neutron matter studies. As\-sum\-ing no
strong phase tran\-si\-tions at lower densities, we expect the minimum
radius limit, to neutron stars, which primarily depends on $p_1$, from
neutron matter constraints to be significantly larger than the
absolute minimum limit, $~8.1$ km~\cite{Lattimer12} established from
the ``maximally compact'' EOS of Ref.~\cite{Koranda97} and an observed
$2~\Msun$ neutron star. $p_1$ will also play an indirect role in
determining $\Mmax$, the maximum neutron star mass, in that stars with
larger radii at intermediate densities typically support larger
masses. On the other hand, we expect that $p_2$ and $p_3$ will play
little role in determining neutron star radii but will play
significant roles in determining $\Mmax$.

\begin{figure}
  \includegraphics[width=8.5cm,angle=180]{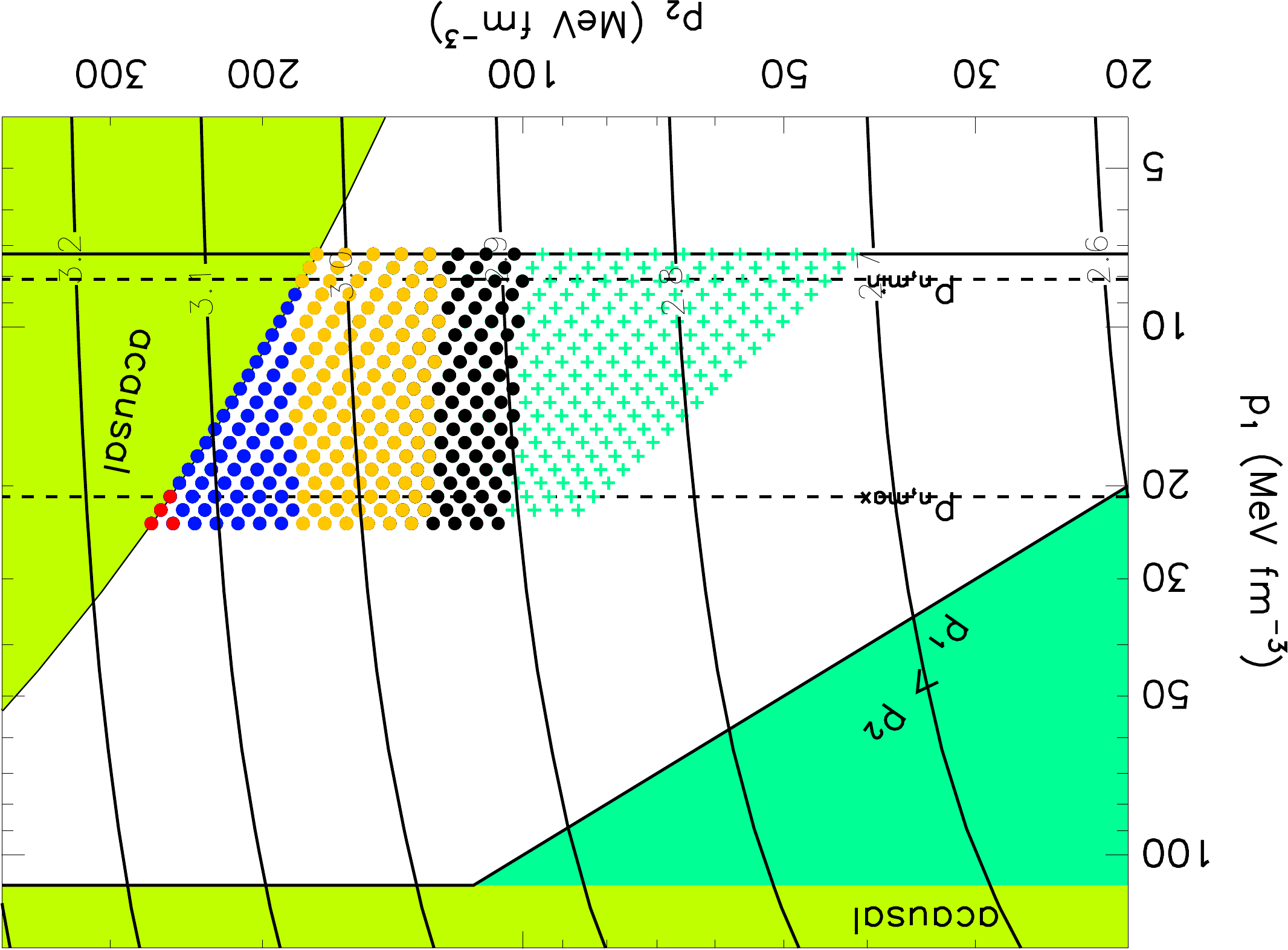}
  \caption{Permitted ranges of $p_1$ and $p_2$. Values ruled out by
    causality (hydrodynamic instability) are indicated by the
    light-(dark-)green shading. Contours of $\log_{10} p_3$ that
    violate acausality are indicated. The realistic minimum and
    maximum values of $p_1$ are shown as horizontal dashed lines, and
    the value $p_{1,\min}=7.56$ MeV fm$^{-3}$ is the horizontal solid
    line. Solid dots show parameter values permitting maximum masses,
    respectively, of $1.97~\Msun$ (black), $2.10~\Msun$ (orange),
    $2.30~\Msun$ (blue) and $2.50~\Msun$ (red). Green crosses show
    parameter values that lead to acausal configurations or those with
    $\Mmax<1.97~\Msun$.}
  \label{fig:p1p2}
\end{figure}

\begin{figure}
  \includegraphics[width=8.5cm,angle=180]{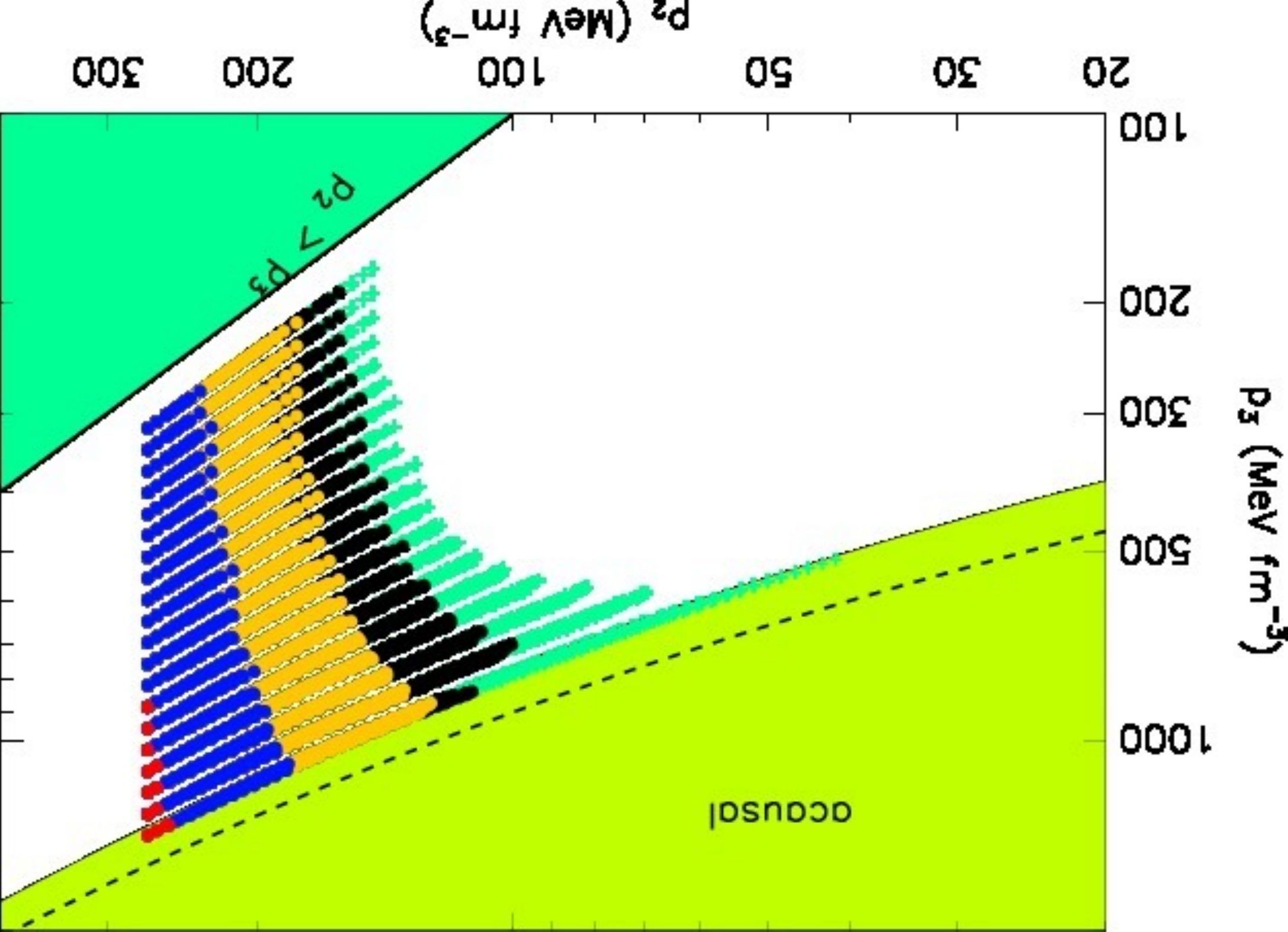}
  \caption{Permitted ranges of $p_2$ and $p_3$. Those values ruled out
    by acausality (hydrodynamic instability) are indicated by the
    light-(dark-)green shading. The acausal region depends slightly on
    the assumed value of $p_1$: the dashed line is for $p_{1,\max}$; the
    edge of the shaded region is for $p_{1,\min}=7.56$ MeV fm$^{-3}$.
    Symbols are described in the caption for Fig. \ref{fig:p1p2}.}
  \label{fig:p2p3}
\end{figure}

Hy\-dro\-dy\-nam\-ic sta\-bi\-li\-ty, caus\-a\-li\-ty, and ob\-served
neu\-tron star masses im\-pose im\-por\-tant con\-straints on the
choi\-ces of pa\-ra\-me\-ters. In this scheme the sound speed
increases monotonically with density within each polytrope, so that
causality within regions 1 to 3 requires
\begin{equation}
  \frac{c_{s,i}^2}{c^2}=\left(\frac{\partial p}{\partial\epsilon}
  \right)_i=\frac{\gamma_{i,\max}p_{i,\max}}{\epsilon_{i,\max}+p_{i,\max}}
  \le 1,
  \label{eq:caus}
\end{equation}
which gives implicit equations for $p_{i,\max}$ since
\begin{equation}
  \epsilon=\frac{p}{\gamma_i-1}+\left(\epsilon_{i-1}-
  \frac{p_{i-1}}{\gamma_i-1}\right)
  \left(\frac{p}{ p_{i-1}}\right)^{1/\gamma_i}.
  \label{eq:emax}
\end{equation}
Therefore, $p_{1,\max}$ depends upon $p_0$ and $\epsilon_0$, and
$p_{2,\max}$ depends upon $p_1$ and $\epsilon_1$, but $\epsilon_1$
depends on $p_0$. In general, $p_{1,\max}$ is so much larger than the
realistic ranges of $p_1$ established from neutron matter studies that
the causality condition is not important to our studies
($p_{1,\max}\simeq113.9$ MeV fm$^{-3}$ from causality considerations (Fig. \ref{fig:p1p2})).
$p_{3,\max}$ depends upon $p_2$ and $\epsilon_2$, and therefore also
upon $p_1$. However, the dependence upon $p_1$ is weak, as shown in
Fig. \ref{fig:p2p3}. In addition, the central density of the star is,
in many cases, larger than $n_3$, in which case the limiting value
$p_{3,\max}$ is smaller than the limit given by Eq. (\ref{eq:caus}).
The actual limit must be found numerically from TOV integrations of
the star's structure requiring that the maximum sound speed at the
center of the maximum mass star be smaller than $c$.

Minimum values for $p_1$, $p_2$ and/or $p_3$ also exist in order to
satisfy hydrodynamic stability, which requires that $p_i>p_{i-1}$.
This is obviously satisfied by any realistic value of $p_1$.  Parameter ranges allowed by hydrodynamical stability and causality are
portrayed in Figs. \ref{fig:p1p2} and \ref{fig:p2p3} as the white
regions. For $p_2$
and $p_3$, more restrictive minima can result from the constraint that
the maximum mass exceeds the largest well-measured neutron star mass,
which we take to be $1.97~\Msun$, the $1\sigma$ lower limit to the
measured mass of PSR J0548+0432~\cite{Antoniadis13}. (Note that there
is no minimum value for $p_1$ based on this condition, due to the
presence of the polytropic regions 2 and 3.) These lower limits also
must be found numerically from TOV integrations, which indicates that
the effective lower limit to $p_2$ is approximately 100 MeV fm$^{-3}$
for virtually all realistic choices of $p_1$ (Fig. \ref{fig:p1p2}).

The result of each TOV integration with a different EOS (i.e., different combinations of $p_1, p_2$ and $p_3$) is indicated
by a symbol in Figs.~\ref{fig:p1p2} and \ref{fig:p2p3} (many parameter
combinations yield nearly identical configurations and cannot be
distinguished). Solid circles show parameters that support
causal configurations with, respectively, $\Mmax=1.97~\Msun$ (black),
$2.10~\Msun$ (orange), $2.30~\Msun$ (blue) and $2.5~\Msun$ (red).
Parameters that yield acausal configurations, those in
which the sound speed at the center exceeds $c$, or configurations
incapable of supporting at least $1.97~\Msun$, are shown as teal
crosses. It is clear that causal configurations capable of supporting
$\Mmax=1.97~\Msun$ must have $p_2\simge100$ MeV fm$^{-3}$, and if
$\Mmax=2.1~\Msun$, $p_2\simge125$ MeV fm$^{-3}$. On the other hand,
the specific value of $p_3$ plays relatively little role as long as
$p_2<p_3<p_{3,\max}$ lead to causal configurations of the required
mass.

We note that the parameter boundaries due to causality we find are in
disagreement with the results of Ref.~\cite{Ozel15a} in spite of the
fact that the parameterized EOSs in the two studies are identical. For
example, when it is assumed that $\log_{10} p_1=1$,
Ref.~\cite{Ozel15a} indicate that $\log_{10}p_{2,\max}\simeq2.6$,
while we find $\log_{10}p_{2,\max}\simeq2.3$; when it is assumed that
$\log_{10}p_2=1.6$, Ref.~\cite{Ozel15a} indicates that
$\log_{10}p_{3,\max}\simeq3.0$, while we find
$\log_{10}p_{3,\max}\simeq 2.7$. We also note that the solution
preferred by the Bayesian analysis of Ref.~\cite{Ozel15a} has optimum
parameters $p_2$ and $p_3$ that lie very near the acausal boundary,
and that their parameter region with finite likelihoods extends beyond
this boundary.

\subsection{Alternative Models for High-Density Matter}
\label{s:alt}

Choosing different models can be thought of as choosing different
prior distributions, as discussed in Ref.~\cite{Lattimer14co}. Thus,
in order to understand how our results vary with different prior
distributions, we also employ two models from Ref.~\cite{Steiner15un}.
The crust EOS is described in the same way using the results of
Ref.~\cite{Baym71} but is supplemented with the results of
Ref.~\cite{Negele73} at the highest densities in the crust. From $n_0$
to $n_s$, the GCR EOS is used (see above). At higher densities, either
Model A or Model C from Ref.~\cite{Steiner13tn} is used. Model A is
built on piecewise polytropes and gives very similar results to our
first model described above. Model C is constructed using line
segments in the $P$-$\varepsilon$ plane and prefers stronger phase
transitions (this model is called ``qmc\_fixp'' in Ref.~\cite{bamr}).
In particular, Model C allows strong phase transitions just above
$n_s$. This can make a significant difference in the results, as shown
below. It is clear that exotic matter cannot appear at $n_s$ since
laboratory nuclei are known to consist only of neutrons and protons.
The minimum possible density at which a phase transition may appear,
however, is not well-determined from either theory or experiment.

We can also investigate the observed differences between Models A and
C within the piecewise polytropic method described in Section
\ref{s:piece} by explicitly creating a first-order phase transition
between the fiducial densities $n_1$ and $n_2$ and allowing those
densities to vary. One requires that $p_1=p_2$ and $\mu_1=\mu_2$,
where $\mu=(\epsilon+p)/n$ is the chemical potential. The maximum
effect of the phase transition results if the EOS above $n_2$ is given
by the causal limit, $p=\epsilon-\epsilon_2+p_2$. In this
causality-limited case, one can show that the phase transition
strengths $\epsilon_2/\epsilon_1$ and $n_2/n_1$ are related by:
\begin{equation}
  {\epsilon_2\over\epsilon_1}={p_1\over\epsilon_1}\left({n_2\over n_1}-
  1\right)+{n_2\over n_1}.
\label{eq:phase}\end{equation}

\section{Results}

\subsection{Masses, Radii, and the EOS}

For the polytropic EOS from section~\ref{s:piece}, TOV integrations
were computed with a grid of values for $p_1$, $p_2$ and $p_3$.
Generally, we assume $p_1$ to be limited by the range described above
for realistic neutron matter studies. However, we slightly extended
both the lower and upper limits of $p_1$ to be conservative. Following
Ref.~\cite{Ozel15a}, we choose the value $p_{1,\min}=7.56$ MeV
fm$^{-3}$, which proves crucial to the computation of the minimum
realistic neutron star radius. We increased the upper limit to be
about 1.6 times larger than $p_{n,\max}$, which proves crucial in
establishing a maximum realistic neutron star radius. Values of $p_2$
were chosen to be smoothly distributed in $\log_{10} p_2$ between
$\log_{10} p_{2,\min} = 1.6$ MeV fm$^{-3}$ (to avoid unnecessary
computations of configurations with $\Mmax\simle1.90~\Msun$) and
$p_{2,\max}$ (determined from causality considerations). Values for
$p_3$ were taken to be smoothly distributed in $\log_{10} p_3$ between
$p_2$ (hydrodynamic stability) and $p_{3,\max}$ (causality up to
$n_3$).

\begin{figure}
  \includegraphics[width=8.5cm,angle=180]{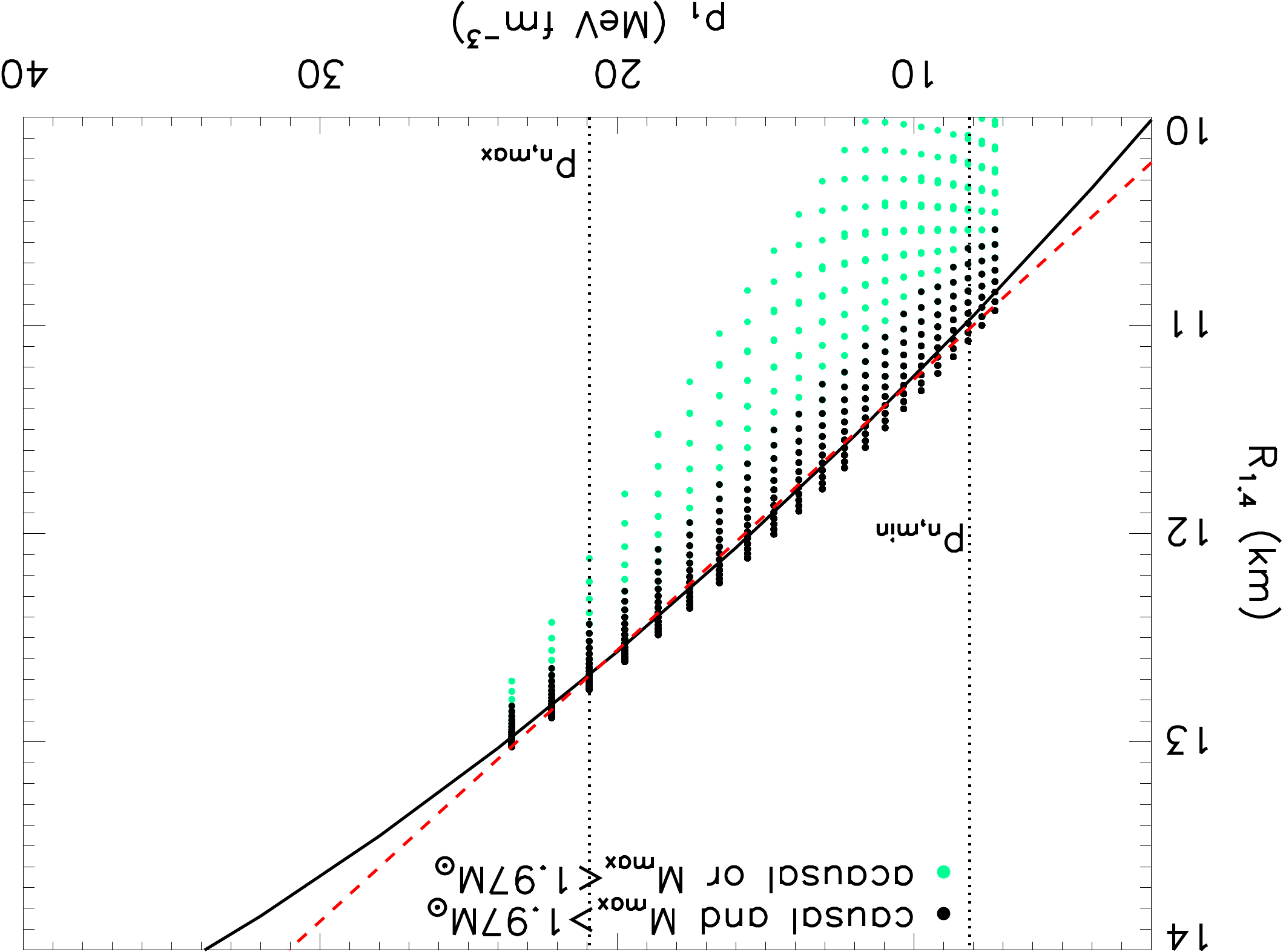}
  \caption{The correlation between radii of $1.4~\Msun$ stars,
    $R_{1.4}$, and $p_1$. Parameters producing causal configurations
    capable of supporting $1.97~\Msun$ are indicated as black circles;
    all others are indicated by green circles. The solid (dashed)
    lines indicate quadratic (linear) fits to the black
    circles.}
  \label{fig:rad}
\end{figure}

\begin{figure}
  \includegraphics[width=8.5cm]{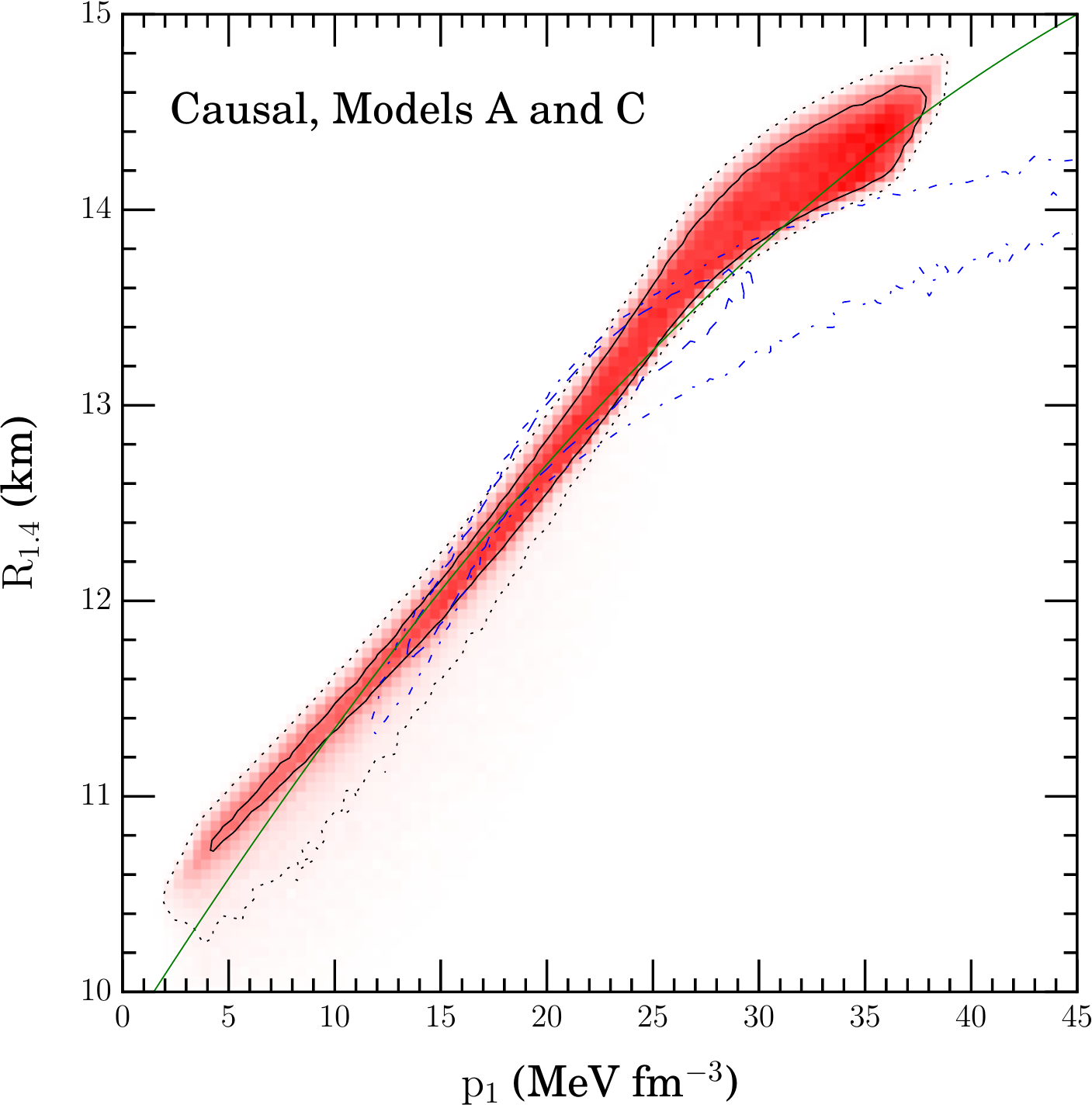}
  \caption{Correlation between $R_{1.4}$ and the pressure, $p_1$ for Model A
    (dashed and dashed-dotted contours give 68\% and 95\% confidence
    regions, respectively) and Model C (solid and dotted contours and
    shaded histogram). Acausal EOSs and those which have maximum
    masses less than $1.97~\mathrm{M}_{\odot}$ have been removed.
    The correlation from Eq.~\ref{eq:r14p1} is also plotted.}
  \label{fig:R14P1p85both}
\end{figure}

We also use Model C EOS as described in section~\ref{s:alt}. The
calculation proceeds as for the polytropic EOS described above, except
that the results for $>10^5$ Monte Carlo realizations are histogrammed
and binned (using Ref.~\cite{o2scl}). In effect, because Model C
prefers stronger phase tran\-si\-tions, com\-pa\-ring with the
po\-ly\-tro\-pic EOS com\-pares two different prior assumptions. The
correlation between the radius of 1.4~$\mathrm{M}_{\odot}$ and the
pressure, $p_1$ is plotted for the polytropic EOS in
Fig.~\ref{fig:rad}, and for Model A and Model C in
Fig.~\ref{fig:R14P1p85both}. As expected, a large degree of
correlation exists between the radii of $1.4~\Msun$ configurations,
$R_{1.4}$, and $p_1$ in either case. Eliminating acausal
configurations, or configurations incapable of supporting
$1.97~\Msun$, which are shown in Fig.~\ref{fig:rad} as green circles,
the correlation becomes more robust. The correlation between $p_1$ and
$R_{1.4}$ is best described with a quadratic contribution, as can be
seen in Figs. \ref{fig:rad} and \ref{fig:R14P1p85both}. The
correlation between $p_1$ and $R_{1.4}$ is
\begin{align}
R_{1.4}&=9.68+0.168p_1-0.00120p_1^2{\rm~km},
\label{eq:r14p1}
\end{align}
where $p_1$ is in units of MeV fm$^{-3}$. 

\begin{figure}
  \includegraphics[width=8.5cm,angle=180]{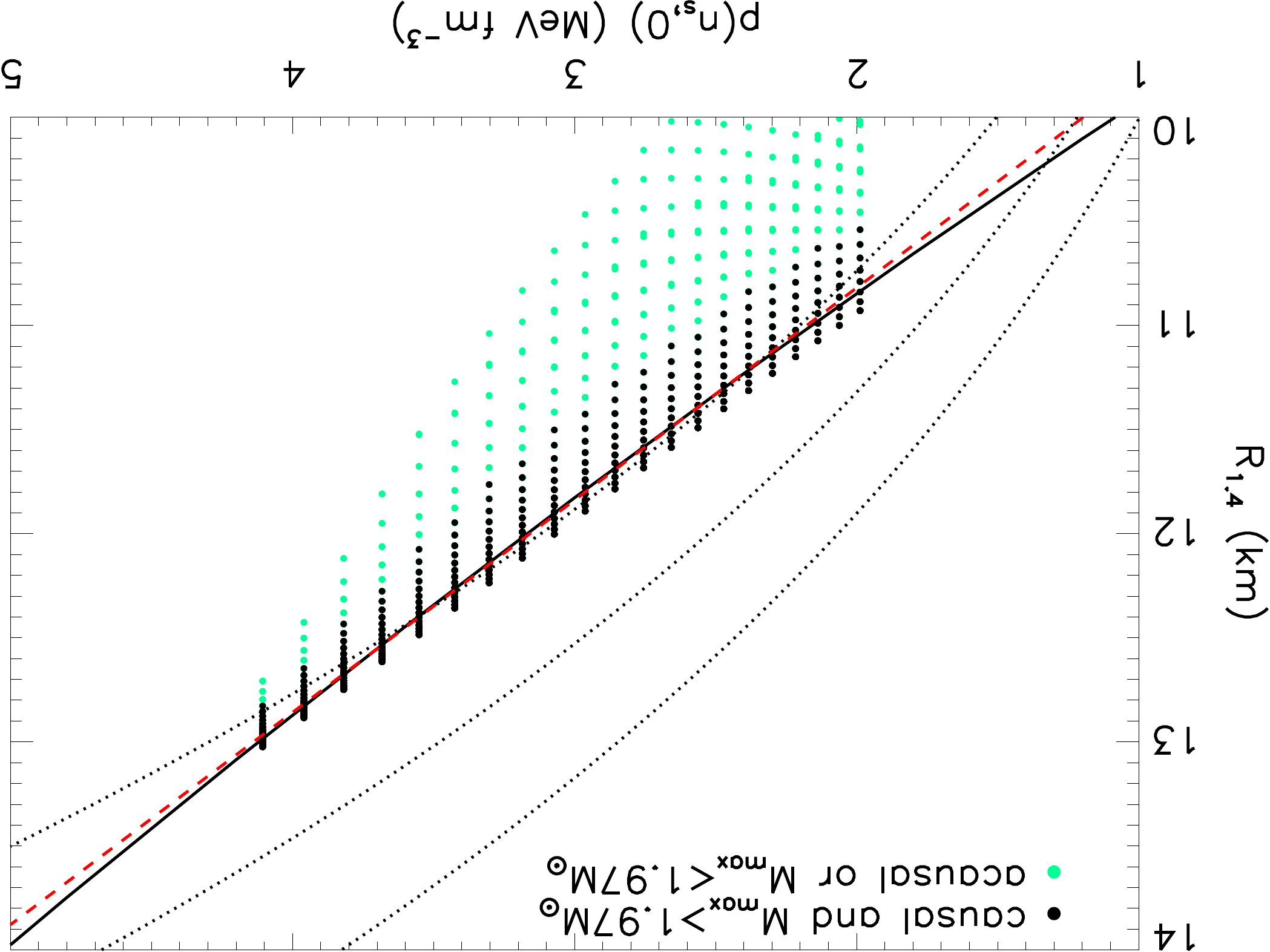}
  \caption{The same as Fig. \ref{fig:rad} except showing the
    correlation between $R_{1.4}$ and $p(n_s,0)$. The dotted curves
    show the correlation determined by Ref. \cite{Lattimer13} with
    $1\sigma$ errors.}
\label{fig:rad1}
\end{figure}

\begin{figure}
  \includegraphics[width=8.5cm]{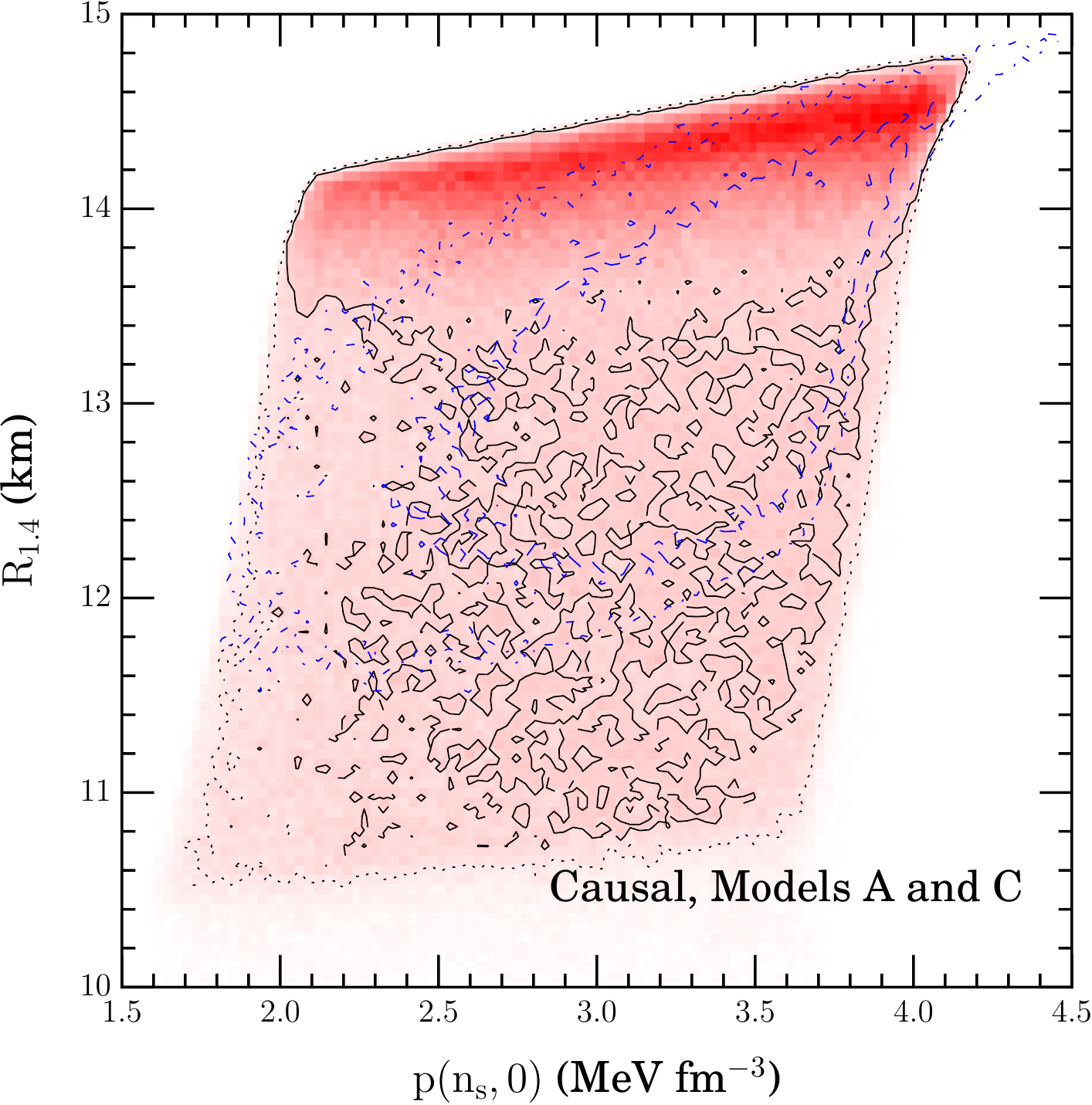}
  \caption{The correlation between $R_{1.4}$ and the pressure,
    $p(n_s)$, formatted as in Fig.~\ref{fig:R14P1p85both}. Model A,
    which is based on polytropes, does show a slight correlation.}
  \label{fig:R14P1both}
\end{figure}

Ref.~\cite{Lattimer01} had shown the existence of a phenomenological
relation between $R_{1.4}$ and $p(n_s,x_\beta)$, which was later
modified~\cite{Lattimer13} to consider only EOSs capable of supporting
$2~\Msun$:
\begin{equation}
R_{1.4}=(9.52\pm0.49)[p(n_s,x_\beta)/({\rm MeV~fm}^{-3})]^{1/4}{\rm~km}.
\label{eq:lp}
\end{equation}
The polytrope results are consistent with this relation (Fig.
\ref{fig:rad1}), where we approximated $p(n_s,x_\beta)\simeq
p(n_s,0)=3L/n_0$ from the piecewise polytropic EOSs, as can readily be
expected given the high degree of correlation between $p_1$ and
$p(n_s,0)$ or $L$ (Fig. \ref{fig:p1l}). The tendency for our predicted
radii to be slightly smaller than those predicted by Eq. (\ref{eq:lp})
is due to the fact that both the lower and upper limits to $p_1$ are
smaller than for the range of the EOS samples considered by
Refs.~\cite{Lattimer10} and \cite{Lattimer13}. However, this
correlation is strongly sensitive to the prior assumptions, i.e. the
possible presence of a phase transition just above the saturation
density. This is demonstrated by the stark difference between the
polytrope-based models and Model C (compare Figs.~\ref{fig:rad1} and
\ref{fig:R14P1both}.)

The stark difference in permitted values of $R_{1.4}$ is readily
understood if one permits strong phase transitions and a high-density
causality-limited EOS in the piecewise polytrope model by permitting
$n_1$ and $n_2$ to vary, as described in Sec. \ref{s:alt}. There it
was shown, Eq. (\ref{eq:phase}), that the strength of the phase
transition is described by $n_2/n_1$, or, equivalently, by
$\epsilon_2/\epsilon_1$. The range of permitted values of $R_{1.4}$ is
extremized when $n_1=n_s$. The largest value, $R_{1.4}\simeq14.3$ km,
results when $n_2=n_1=n_s$. The smallest value, $R_{1.4}\simeq8.4$ km,
results when $n_2/n_1\simeq4.2$, and is nearly as small as allowed by
general relativity and causality, 8.15 km \cite{Lattimer12}, in the
case when $\Mmax=1.97M_\odot$. The lower limit to $R_{1.4}$ increases
for larger values of $\Mmax$ (accompanied by a decrease in $n_2/n_1$).
Furthermore, a phase transition is only allowed, assuming
$\Mmax\simle1.97M_\odot$, if $n_2/n_1\simle2$; further decreases in
permitted values of $n_2/n_1$ result for larger values of $\Mmax$.
These conclusions are similar to those of Ref. \cite{Alford15}.

\begin{figure}
  \includegraphics[width=8.5cm,angle=180]{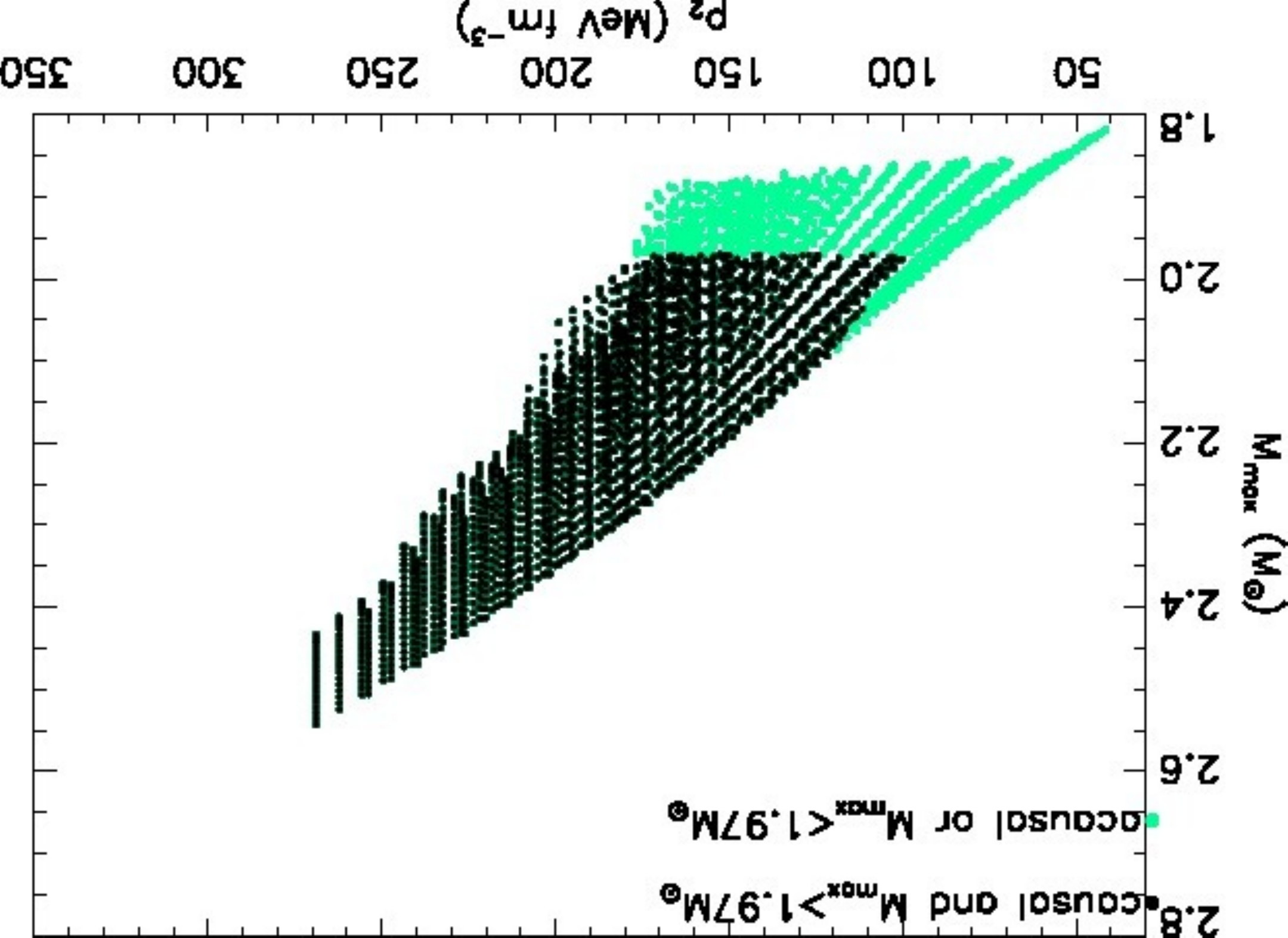}
  \caption{The same as Fig. \ref{fig:rad}, except showing the
    correlation between $\Mmax$ and $p_2$.}
  \label{fig:maxp2}
\end{figure}

\begin{figure}
  \includegraphics[width=8.5cm]{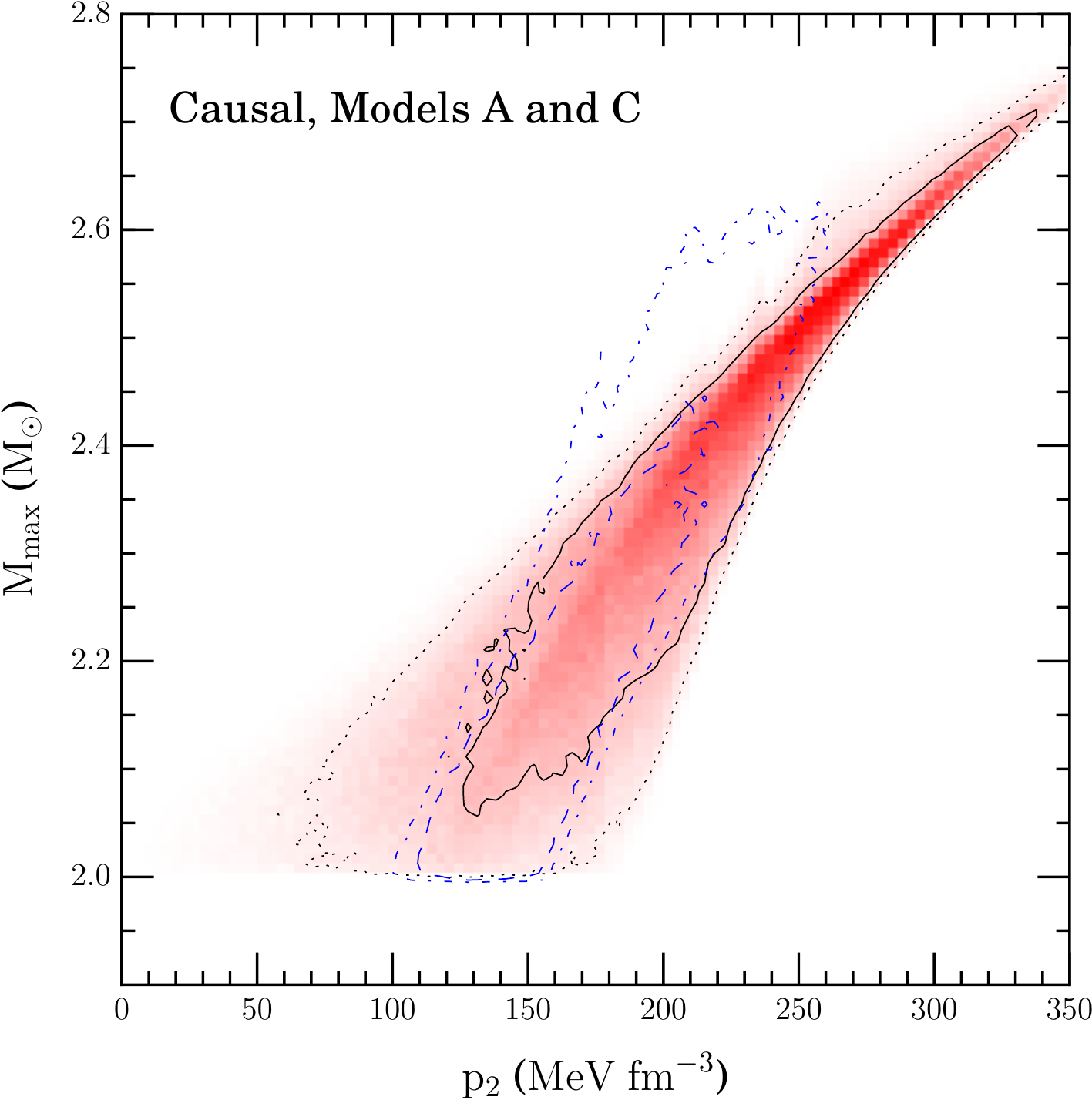}
  \caption{Correlation between $\Mmax$ and the pressure, $p_2$,
    formatted as in Fig.~\ref{fig:R14P1p85both}.}
  \label{fig:P2Mmaxboth}
\end{figure}

There is a lack of correlation between $p_2$ and $R_{1.4}$. Instead, a
high degree of correlation exists between $p_2$ and $\Mmax$ (Figs.
\ref{fig:maxp2} and \ref{fig:P2Mmaxboth}). The presence of this
correlation appears to be relatively prior-independent. Model C, which
allows for stronger phase transitions, also tends to allow EOSs which
have higher pressures at high densities, and thus gives relatively
larger maximum masses. That is, the posterior distribution of $\Mmax$
is strongly prior dependent.

\begin{figure}
  \includegraphics[width=8.5cm]{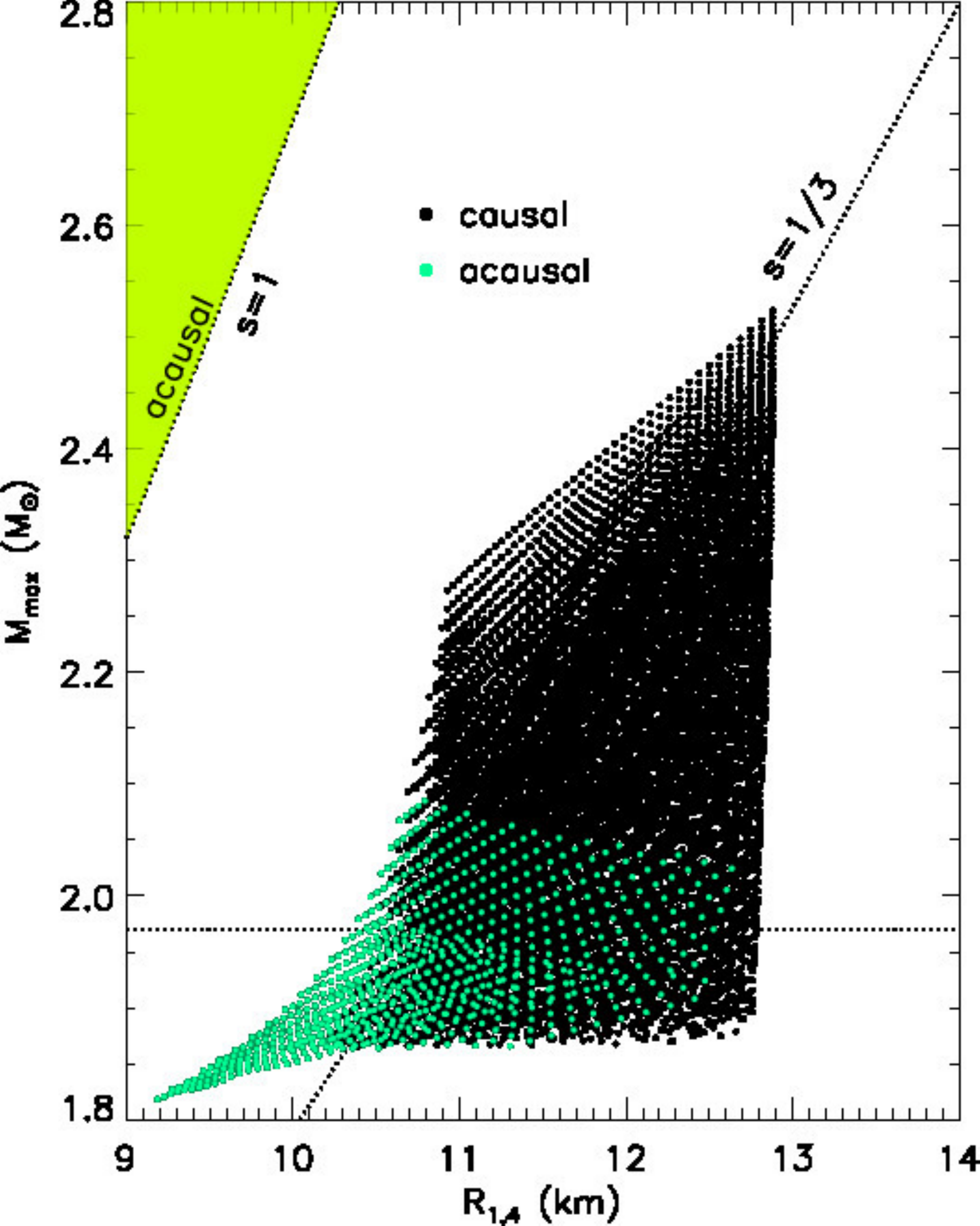}
  \caption{The maximum masses and radii of $1.4~\Msun$ stars predicted
    by piecewise polytropic EOSs. Shown for comparisons are limiting
    radii predicted by the maximally compact EOS with either $s=1$ or
    $s=1/3$. Black circles are causal configurations; green circles
    indicate acausal configurations. }
  \label{fig:max14}
\end{figure}

\begin{figure}
  \includegraphics[width=8.5cm]{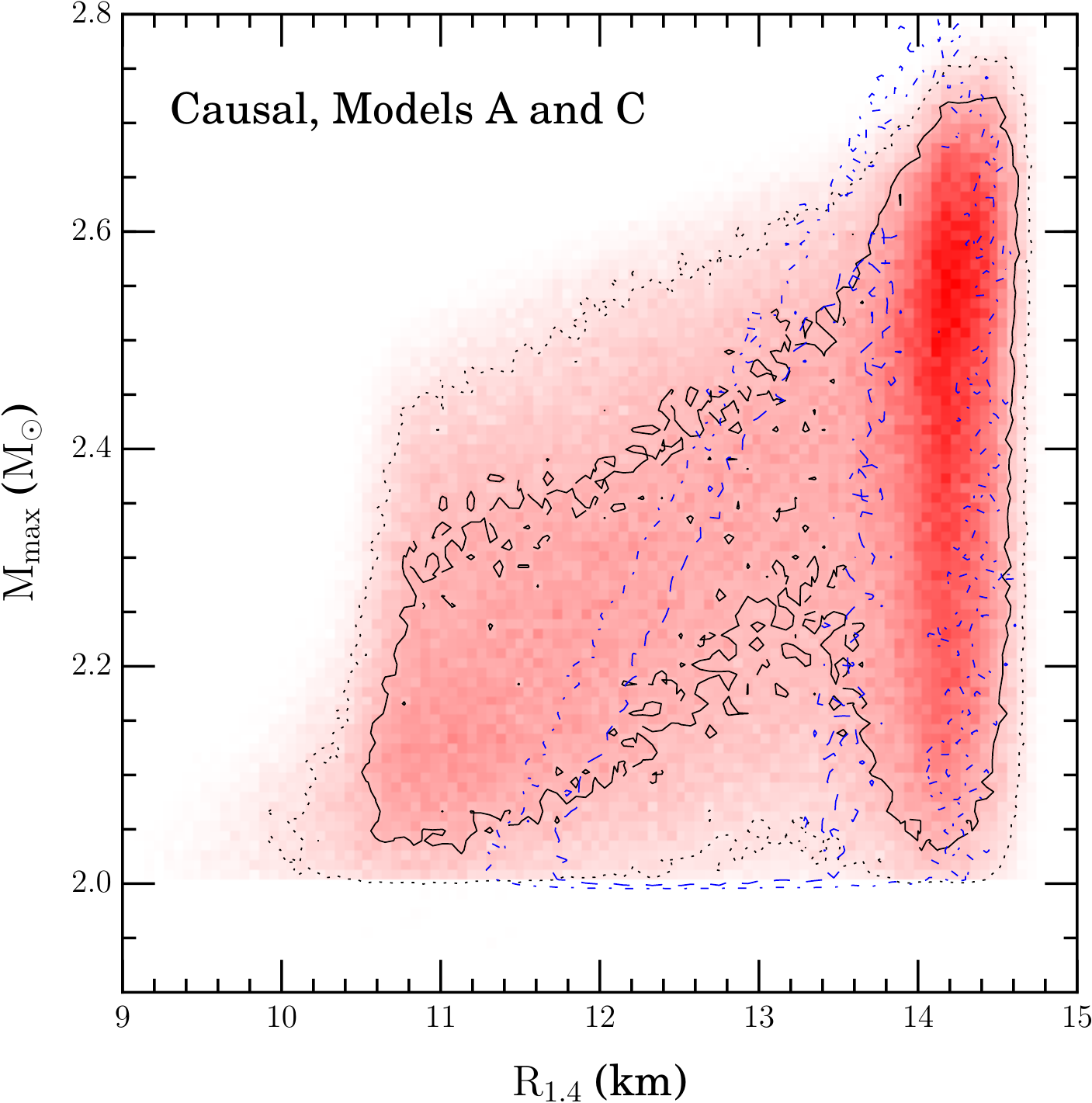}
  \caption{Correlation between $\Mmax$ and the radii of $1.4~\Msun$
    stars, formatted as in Fig.~\ref{fig:R14P1p85both}.}
  \label{fig:R14Mmaxboth}
\end{figure}

Ref.~\cite{Lattimer12} determined the minimum radii of $1.4~\Msun$
stars as a function of the minimum value of $\Mmax$ using the
maximally compact EOS ($s=1$) from Ref.~\cite{Koranda97}. The relation
between $\Mmax$ and $R_{1.4}$ for the piecewise polytropic EOS is
shown in Fig \ref{fig:max14}. Configurations that are causal and
capable of supporting at least $1.97~\Msun$ have $R_{1.4}>10.6$ km
(10.85 km for the minimum value of $p_1$ from neutron matter
calculations). As the maximum mass limit is raised, the lower limit
for $R_{1.4}$ in causal configurations slowly increases. For
$\Mmax>2.4~\Msun$, it is necessary that $R_{1.4}>12$ km and $p_1>14$
MeV fm$^{-3}$. The relation between $\Mmax$ and $R_{1.4}$ for Models A
and C are plotted in Fig.~\ref{fig:R14Mmaxboth}, demonstrating the
dependence on the prior distribution.

Note that causality and the constraint
that $\Mmax>1.97~\Msun$ strongly limits configurations with radii less
then 10 km, as shown in Figs. \ref{fig:max14} and
\ref{fig:R14Mmaxboth}. In the polytrope model, it is important
to emphasize the parameters leading to these acausal configurations
satisfy causality at the boundary points $n_1, n_2$ and $n_3$.
Acausality occurs at densities larger than $n_3$ but below the central
density of maximum mass configurations. Thus, the acausality boundary
shown for $p_3$ as a function of $p_2$ indicated in Fig.
\ref{fig:p2p3} is effectively extended to lower values of $p_3$ at
lower values of $p_2$, and also implies a lower limit to $p_2$ when
combined with the constraint $M_{\mathrm{max}}>1.97M_\odot$.

The minimal radius limit from the maximally compact EOS ($s=1$) is
shown as the solid curve in Fig.~\ref{fig:max14} which excludes the
green shaded region. This boundary approximately represents an extreme
limit because most EOSs have larger radii. It was recently
demonstrated that existence of $2~\Msun$ neutron stars implies that
the sound speed must be larger than
$c_s^2=c^2/3$~\cite{Alford13,Bedaque15sv} (the dotted line in Fig.
\ref{fig:max14}), and most of our parameterized EOSs also exceed this
limit.

\begin{figure}
  \begin{center}
    \includegraphics[width=7cm]{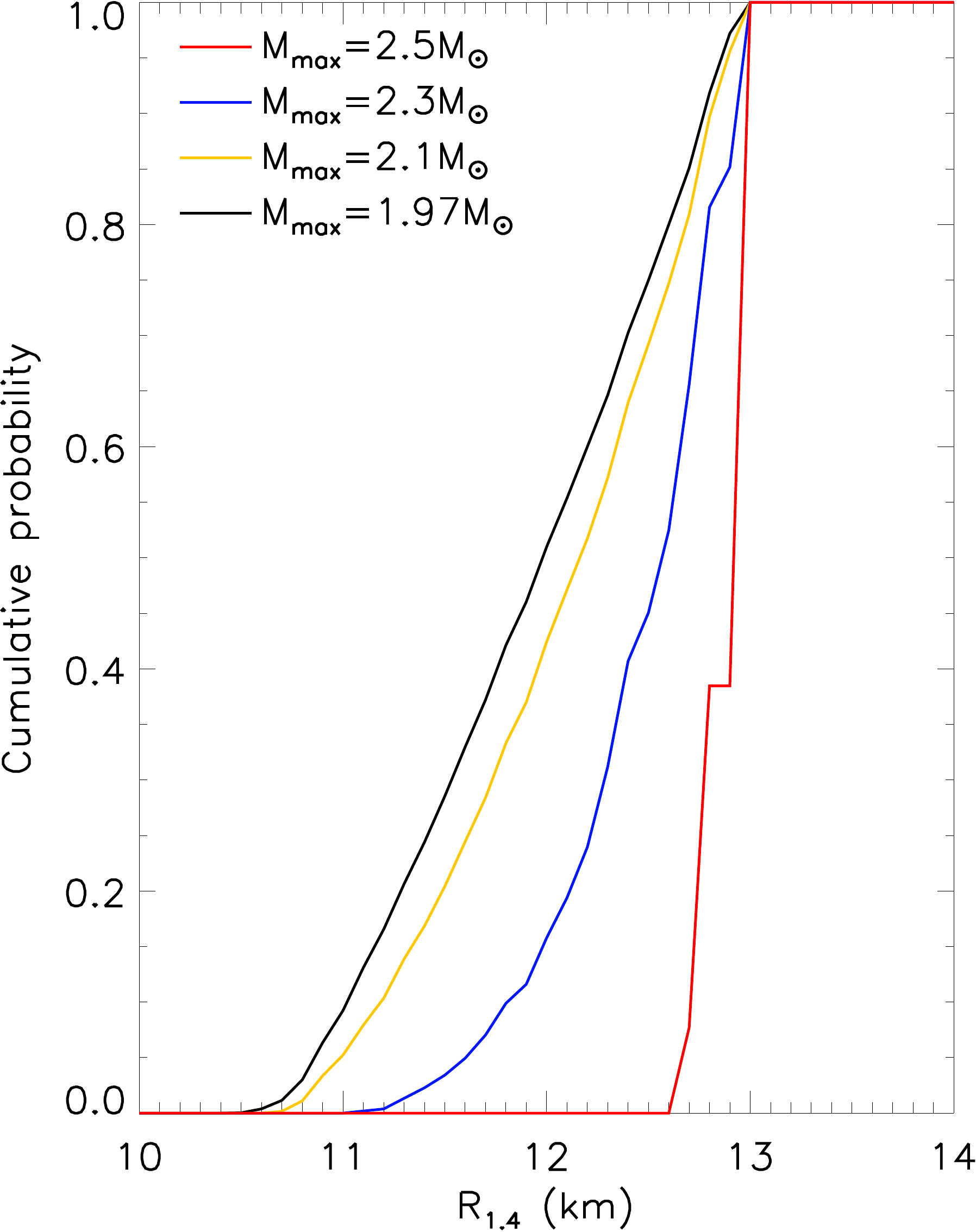}
  \end{center}
  \caption{The cumulative probability of $R_{1.4}$ assuming equal
    likelihoods for parameters $p_1$, $p_2$ and $p_3$ within their
    permitted ranges, in the baryon density polytropic scheme.}
  \label{fig:freq}
\end{figure}

\begin{figure}
  \includegraphics[width=8.5cm]{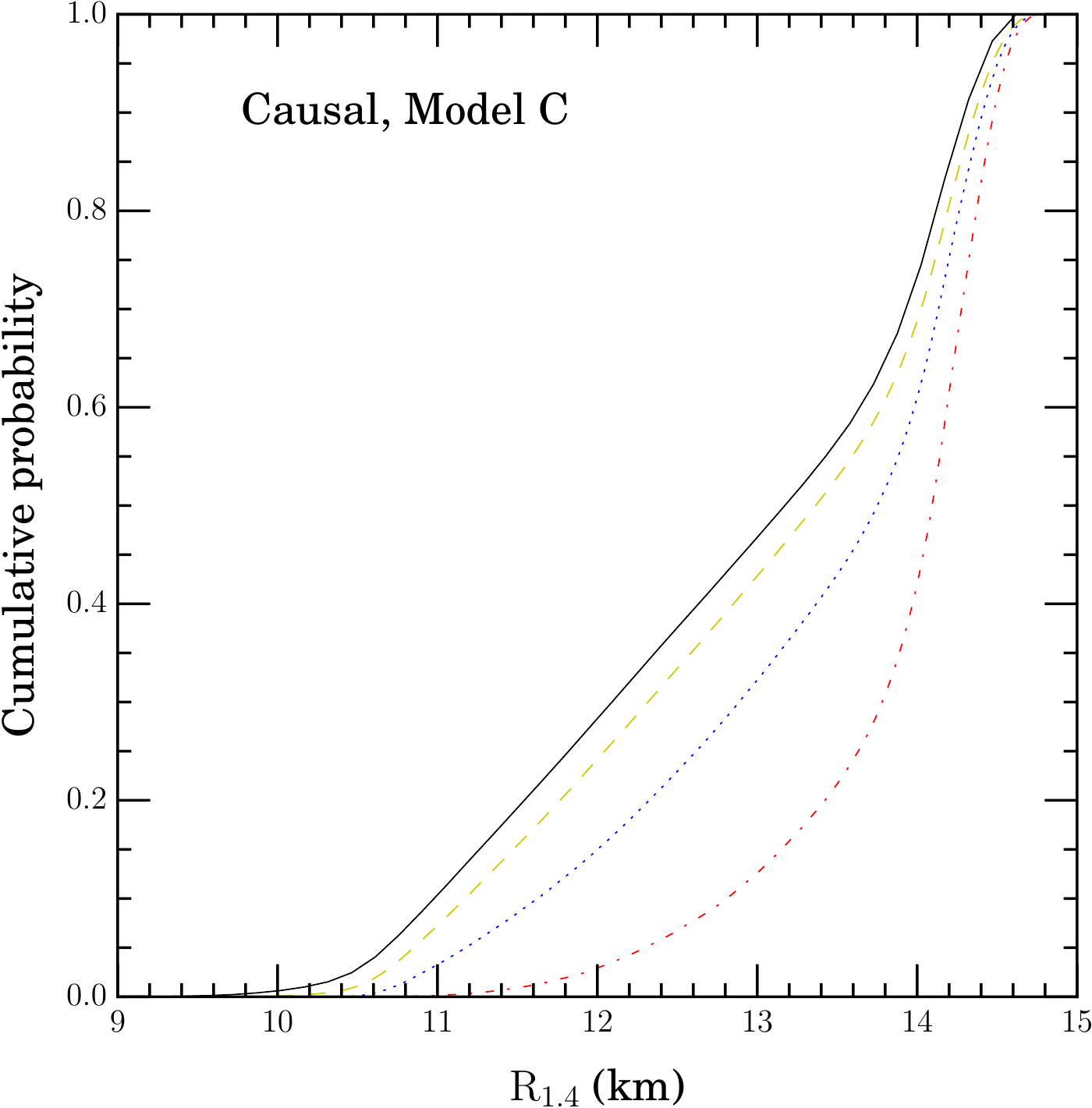}
  \caption{The cumulative probability of $R_{1.4}$ for Model C, a
    prior assumption which allows for stronger phase transitions. The
    (solid, dashed, dotted, and dashed-dotted) lines correspond to
    minimum maximum masses of 1.97, 2.1, 2.3 and 2.5~$\Msun$,
    respectively. Smaller radii are allowed in comparison to
    Fig.~\ref{fig:freq}.}
  \label{fig:freq2}
\end{figure}

In the polytrope model, the realistic upper limit to $p_1$ from
neutron matter theory sets an interesting upper limit to $R_{1.4}$ of
about 13 km. With the present value of $\Mmax\simge1.97~\Msun$,
neutron matter constraints on $p_1$ restrict neutron star radii for
$1.4~\Msun$ stars to lie in the narrow range 11 km~$\simle
R_{1.4}\simle13$ km. It is interesting to examine the frequency
distribution of $R_{1.4}$ among the models that satisfy
$\Mmax>1.97~\Msun$ (Fig. \ref{fig:freq}). Although the minimum radius
is about 10.6 km, assuming that the parameters $p_1, p_2$ and $p_3$
have equal likelihoods within their ranges implies that it is highly
unlikely that $R_{1.4}<11$ km. On the other hand, Model C which allows
strong phase transitions also allows the pressure to increase quickly
above the saturation density. In turn, this allows radii as large as
14.5 km. The corresponding cumulative probability distribution is
given in Fig.~\ref{fig:freq2}. It should be noted however, it is
unclear what physical mechanism would give rise to a strong increase
in the pressure just above the nuclear saturation density. A strongly
repulsive four-neutron force, for example, seems unlikely.

\begin{figure}
  \includegraphics[width=8.5cm,angle=180]{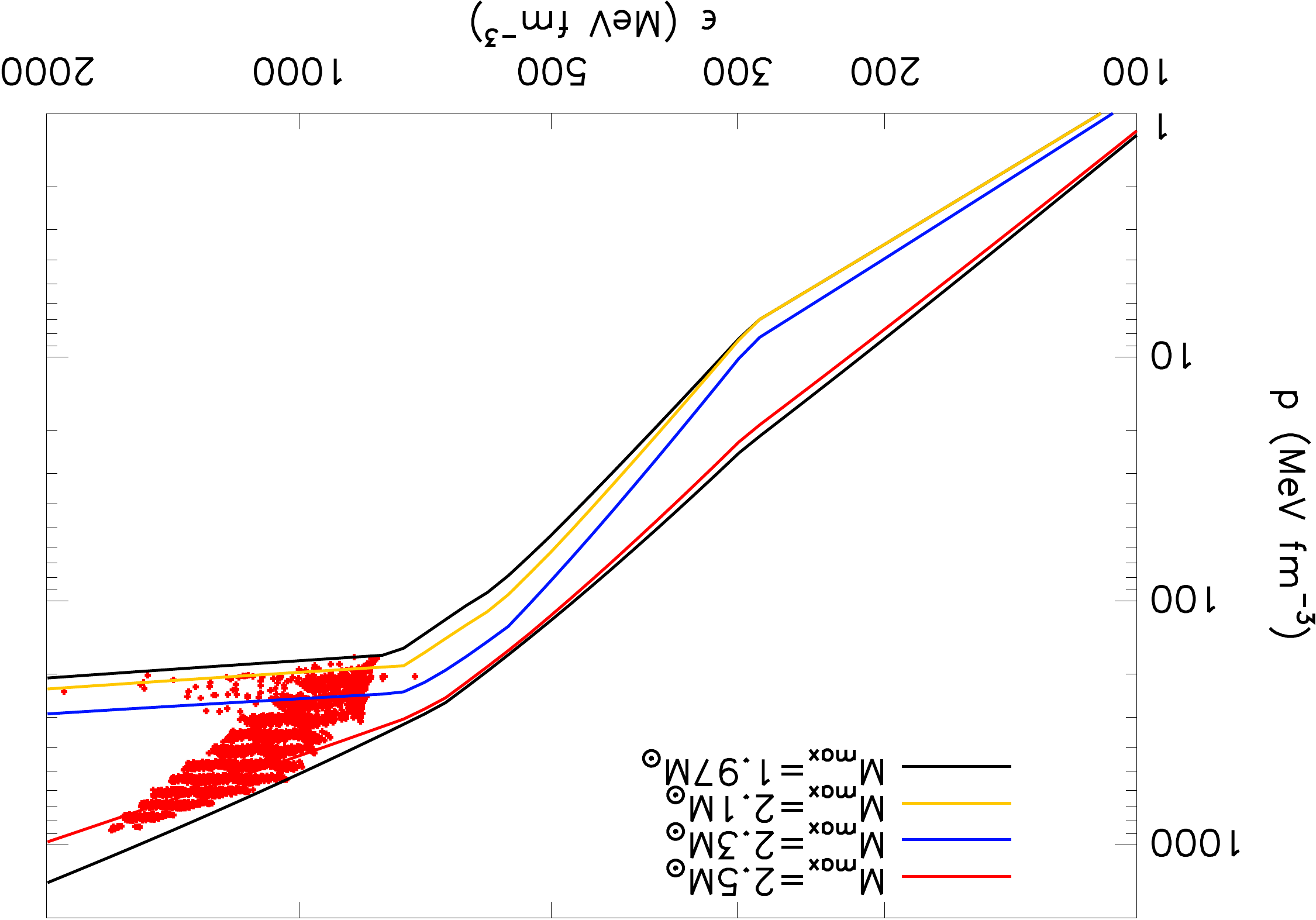}
  \caption{Allowed pressures as a function of energy density permitted
    by the assumed constraints on the low-density EOS, causality, and
    selected values for $\Mmax$. Red crosses indicate the central
    conditions for surviving EOSs. The black, yellow, blue and red lines
    are for $\Mmax=1.97~\Msun$, $2.1~\Msun$, $2.3~\Msun$, and $2.5~\Msun$,
    respectively.}
  \label{fig:pe}
\end{figure}

\begin{figure}
  \includegraphics[width=8.5cm,angle=180]{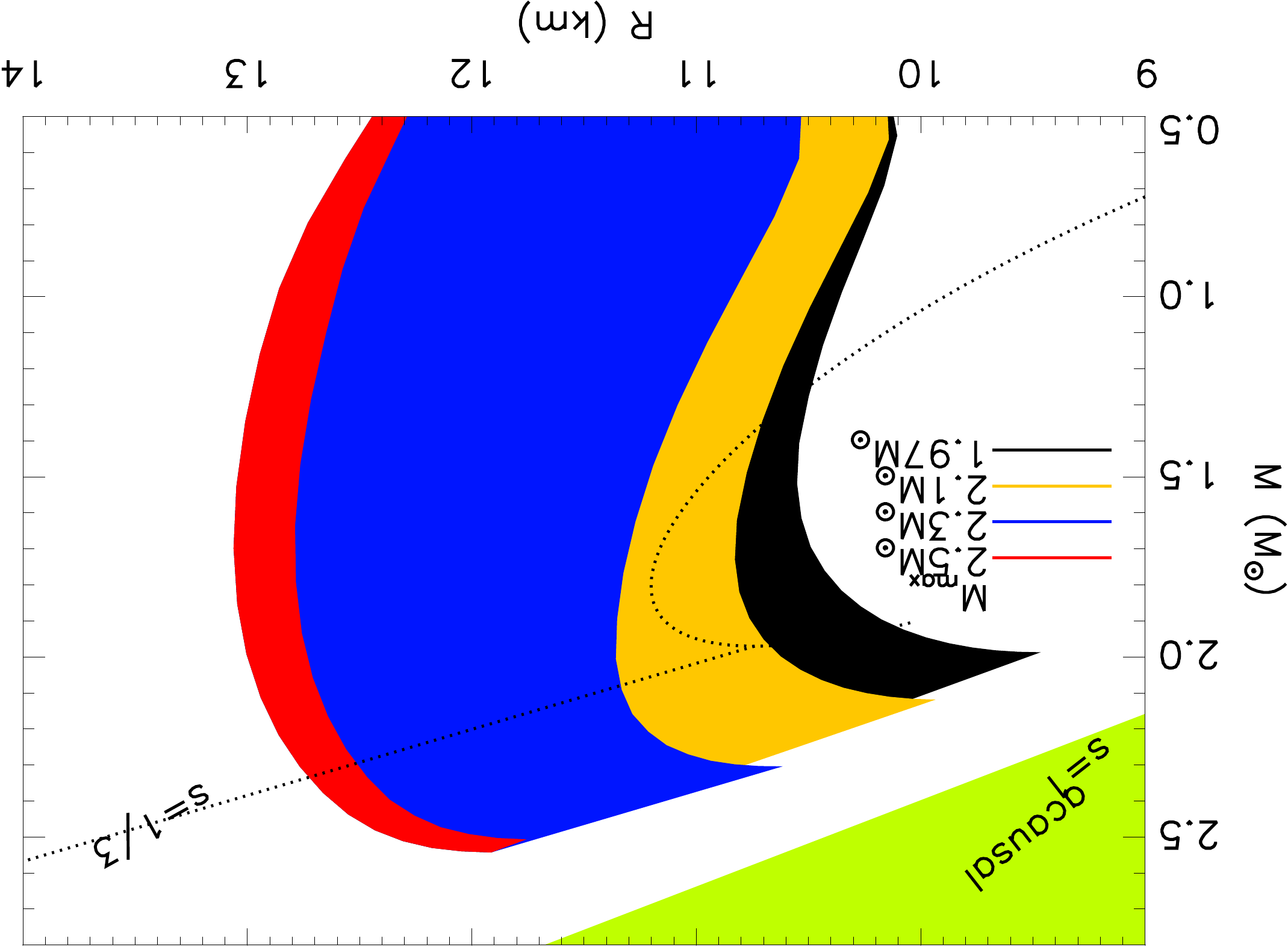}
  \caption{Allowed regions in the $M_R$ plane for selected values of
    $\Mmax$, causality and the assumed constraints for the
    low-density EOS for the polytropic EOS.}
  \label{fig:mr}
\end{figure}

\begin{figure}
  \includegraphics[width=8.5cm]{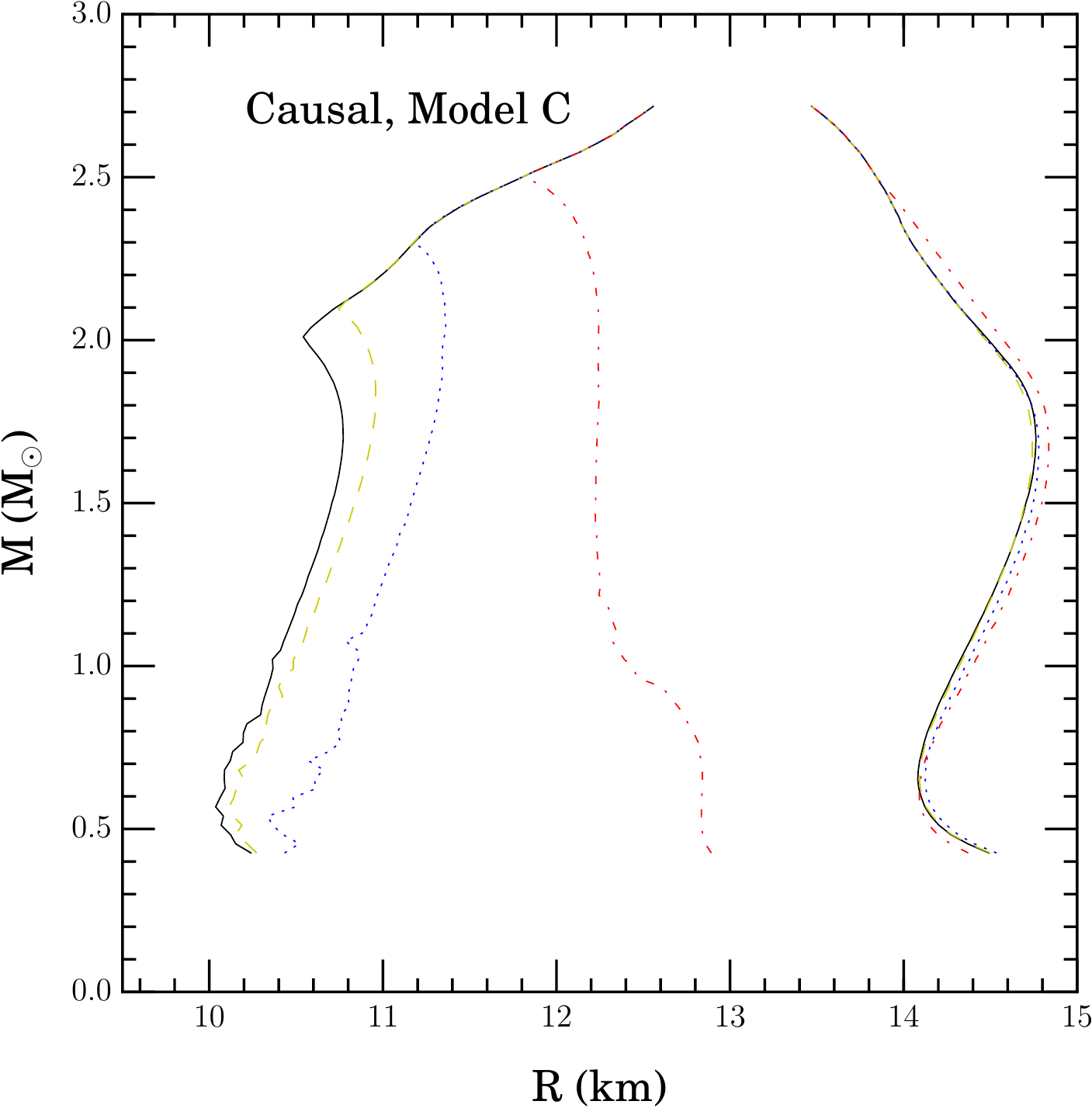}
  \caption{Allowed regions in the $M_R$ plane for selected values of
    $\Mmax$, causality and the assumed constraints for the low-density
    EOS in Model C. The (solid, dashed, dotted, and dashed-dotted)
    lines correspond to minimum maximum masses of 1.97, 2.1, 2.3 and
    2.5~$\Msun$, respectively.}
  \label{fig:mr2}
\end{figure}

The restrictions of causality and large maximum masses severely
restrict the allowed EOSs. Fig. \ref{fig:pe} shows boundaries in the
pressure-energy density plane with different assumptions for $\Mmax$
permitted by causality and the assumed low-density EOS for the crust
and for neutron matter. For $\Mmax=1.97~\Msun$, the maximum
uncertainty in pressure for a given energy density is no larger than a
factor of 3 (which occurs near $n_1$), and is slightly larger than a
factor of 2 near the central densities of maximum mass stars. The
corresponding regions in the mass-radius plane that can be populated
by EOSs satisfying the $\Mmax$, causality and the low-density EOS
constraints are shown in Fig. \ref{fig:mr}. These figures show the
importance of neutron star mass measurements: the larger the minimum
value of $\Mmax$, the more restricted are the ranges of $p(\epsilon)$
and $R(M)$ and the more accurately the EOS can be predicted. The
corresponding regions in the mass-radius phase for Model C are plotted
in Fig.~\ref{fig:mr2}, where larger radii are allowed as discussed
above.

It is interesting to compare our results with those of a recent study
in Ref.~\cite{Chen15}. In that study, various $M$-$R$ curves were
assumed. Beginning with a relativistic mean field EOS (FSU-Garnet)
that predicts $R_{1.4}=13$ km and $\Mmax=2.07~\Msun$, new $M-R$ curves
were generated by arbitrarily translating the original $M-R$ curve for
$M>M_i=0.4~\Msun$ (corresponding to $n_i\simeq1.5n_s$) to smaller
radii by discrete amounts. The EOS corresponding to each
newly-generated $M_R$ curve was deduced by the inversion technique
from Ref.~\cite{Lindblom92}. It was shown that $\Mmax\simeq2~\Msun$
was possible only if $R_{1.4}\simge10.7$ km, a result very similar to
that of the present study if phase transitions are not allowed.
However, it can be argued that the prescription of Ref.~\cite{Chen15}
is model-dependent, being sensitive to the choices of $M_i$ and the
fiducial EOS (FSU-Garnet). In addition, the prescription seems not
consistently applied, as an abrupt reduction of $R$ at $M_i$ implies a
discontinuity in the $M$-$R$ curve and a first-order phase transition
in $p(\epsilon)$. Rather, Ref.~\cite{Chen15} alters the low-density
EOS for $n<n_i$ so that the values of $R_i$ can be reduced from the
original value ($\simeq13$ km) to values as small as 9 km, which
seemingly results in higher pressures for $n\simle0.3n_s$ than the
original EOS. For $n\simeq0.1n_s$, the pressure can be 20--30\% larger
than the original EOS, and the amplification grows with decreasing
density. This is incompatible with our knowledge of the crust's EOS
which allows no such deviations.

\subsection{Universal Relations}

There have been shown to exist several relatively
EOS-in\-de\-pen\-dent re\-la\-tions a\-mong neu\-tron star
ob\-ser\-va\-bles. These may be very useful to reduce degeneracies in
interpretations of observations, including those from gravitational
radiation~\cite{Yagi13}. Ref.~\cite{Lattimer89} found a relation
between the binding energy ${\rm BE}=({\cal N}m_n-M)$, where we set
$G=c=1$, ${\cal N}$ is the number of nucleons in the star, and the
compactness parameter $\beta=M/R$; this was later improved by Ref.
\cite{Lattimer01}, who found
\begin{equation}
{\rm BE}/M\simeq(0.60\pm0.05)\beta(1-\beta/2)^{-1}
\label{eq:bex}
\end{equation}
for a wide variety of EOSs which could support at least $1.65~\Msun$.
Later, Ref.~\cite{Lattimer05s} found another EOS-independent relation
concerning the moment of inertia for EOSs which could support
approximately the same $\Mmax$:
\begin{equation}
I\simeq(0.237\pm0.008)MR^2(1+2.84\beta+18.9\beta^4).
\label{eq:i}
\end{equation}

\begin{figure}
  \includegraphics[width=8.5cm]{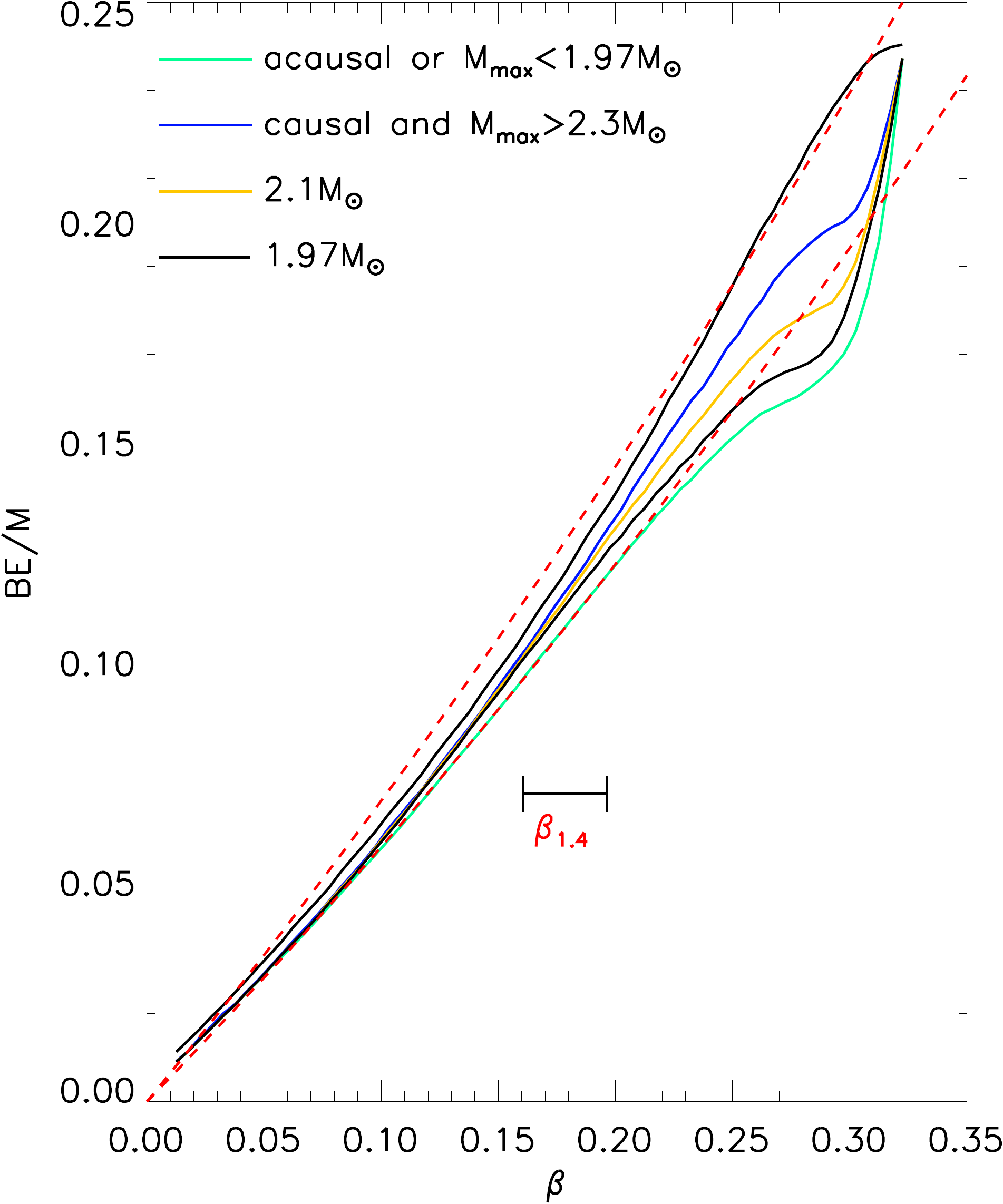}
  \caption{The binding energy as a function of compactness. Solid
    lines show the bounds determined from TOV integrations using the
    full grid of piecewise polytropic EOSs and assuming various values
    for $\Mmax$. Dashed lines indicate Eq. (\ref{eq:bex}). The range
    of compactness for $1.4~\Msun$ stars permitted by causality and
    $\Mmax>1.97~\Msun$ is shown.}
  \label{fig:be}
\end{figure}

\begin{figure}
  \includegraphics[width=8.5cm]{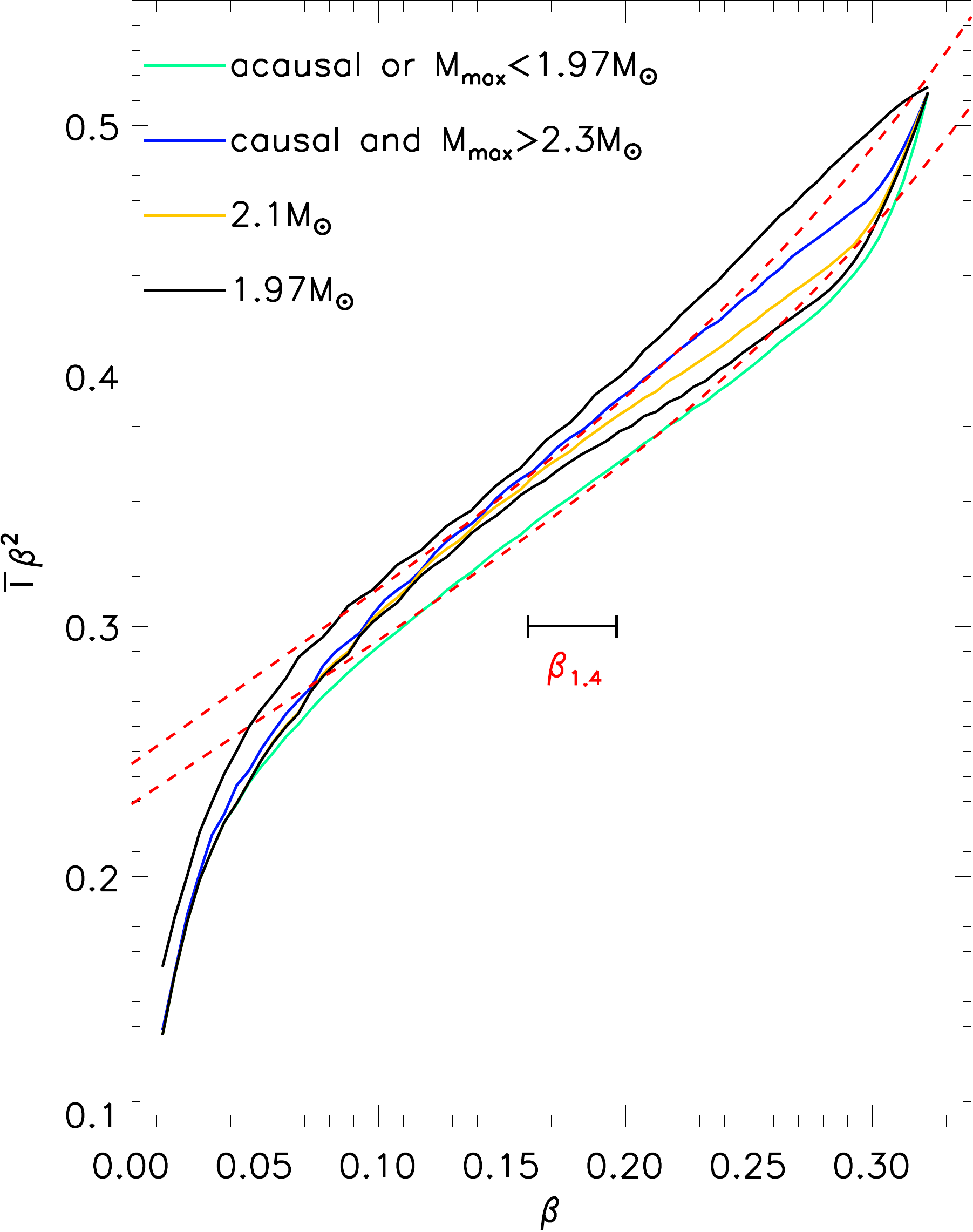}
  \caption{Similar to Fig. \ref{fig:be} but for the moment of inertia
    as a function of compactness. Dashed lines indicate Eq.
    (\ref{eq:i}). }
  \label{fig:i}
\end{figure}

These relations are compared to results from piecewise polytropic EOSs
that satisfy causality and various values of $\Mmax$ in Figs.
\ref{fig:be} and \ref{fig:i}. The results are reasonably well
approximated by Eqs. (\ref{eq:bex}) and (\ref{eq:i}) except for
$\beta\simge0.22$, i.e., close to the maximum masses. We find an
improved approximation for the moment of inertia, showing less
uncertainty, especially for compactnesses typical of $1.4~\Msun$
stars, is given by
\begin{align}
  \frac{I}{ MR^2} & \simeq0.01+
  \left(1.200^{+0.006}_{-0.006}\right)\beta^{1/2}
-0.1839 \beta \\
&-\left(3.735^{+0.095}_{-0.095}\right) \beta^{3/2}+5.278\beta^2.
\label{eq:i1}
\end{align}
The smaller uncertainties result from assuming that
$M_{\mathrm{max}}>1.97~\Msun$. It is apparent, however, from both
Figs. \ref{fig:be} and \ref{fig:i} that should further observations
increase the value of $M_{\mathrm{max}}$ uncertainties in analytic
approximations for BE and $I$ can be substantially reduced.

\begin{figure}
  \includegraphics[width=8.5cm]{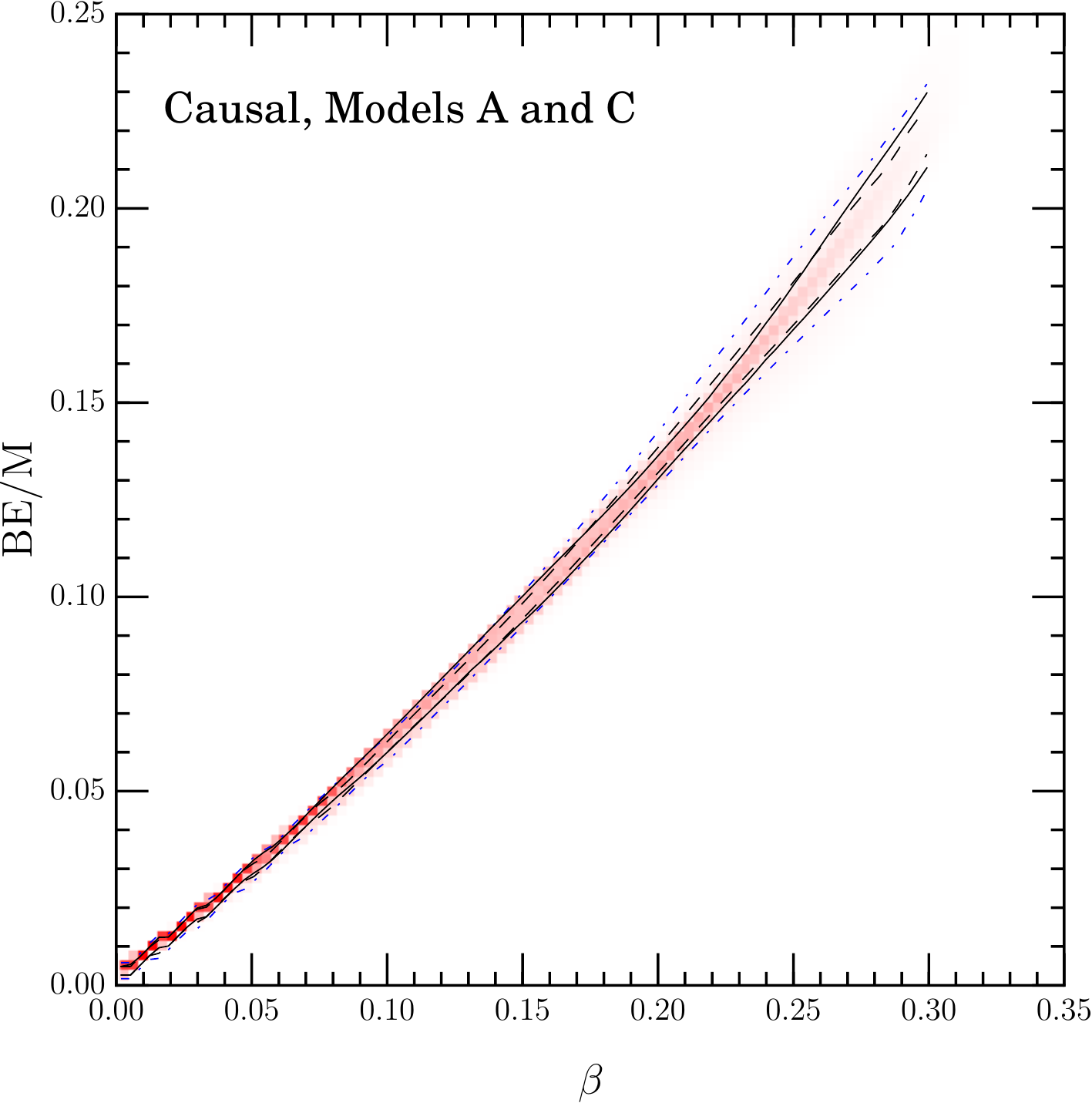}
  \caption{The binding energy as a function of compactness for models
    A and C. The 68\% and 95\% contour lines are indicated with the
    same notation as Fig.~\ref{fig:R14P1p85both} above. Note that the
    correlation is independent of the prior distribution.}
  \label{fig:Bbeta_both}
\end{figure}

\begin{figure}
  \includegraphics[width=8.5cm]{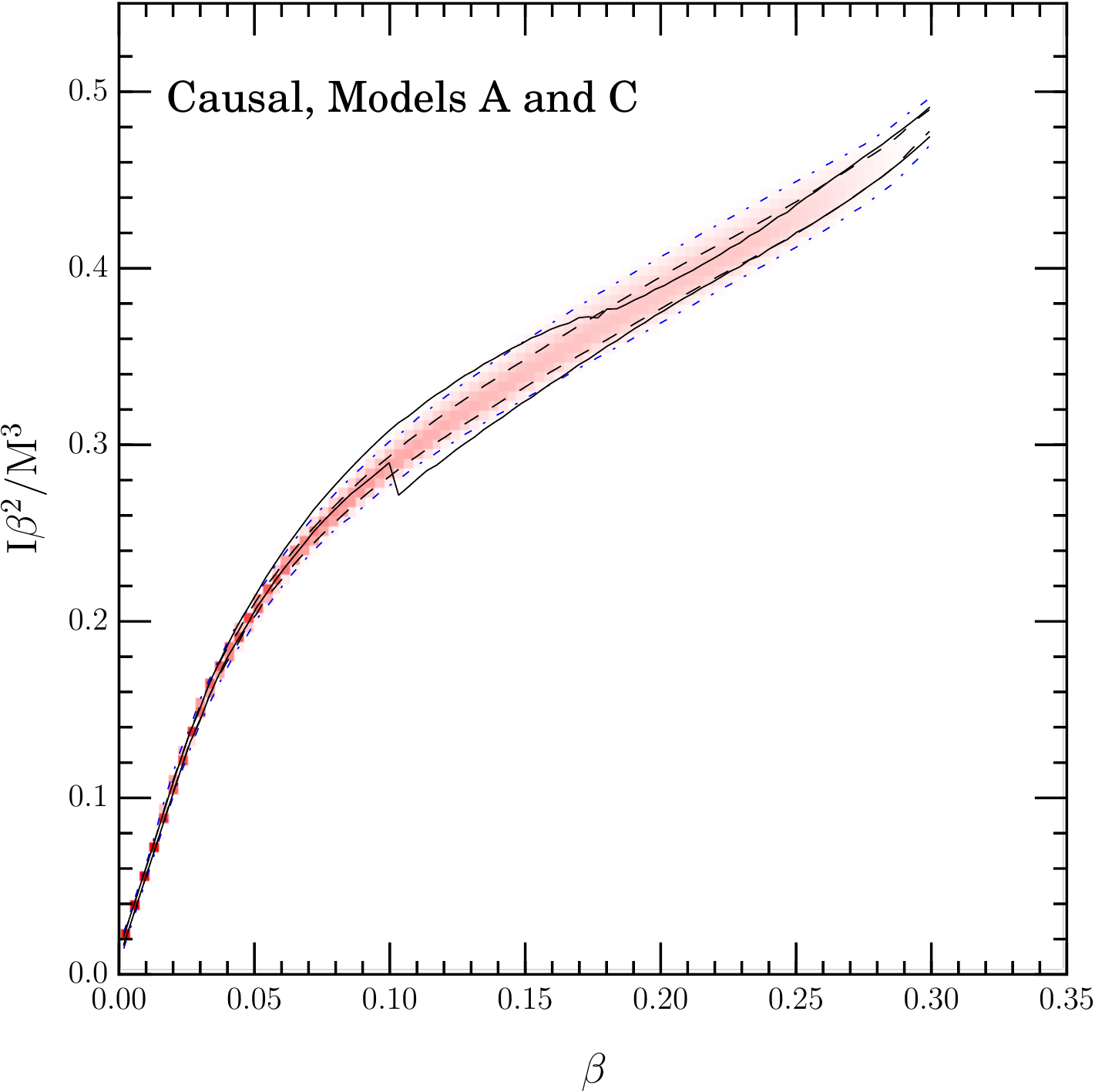}
  \caption{The quantity $I \beta^2/M^3$ as a function of compactness
    for models A and C. The 68\% and 95\% contour lines are indicated
    with the same notation as Fig.~\ref{fig:R14P1p85both} above. Note
    that the correlation is independent of the prior distribution. The
    jump in the contours for Model C at $\beta=0.1$ is a result of
    selecting a new peak in a slightly bimodal distribution. }
  \label{fig:ib2_beta_both}
\end{figure}

\begin{figure}
  \includegraphics[width=8.5cm]{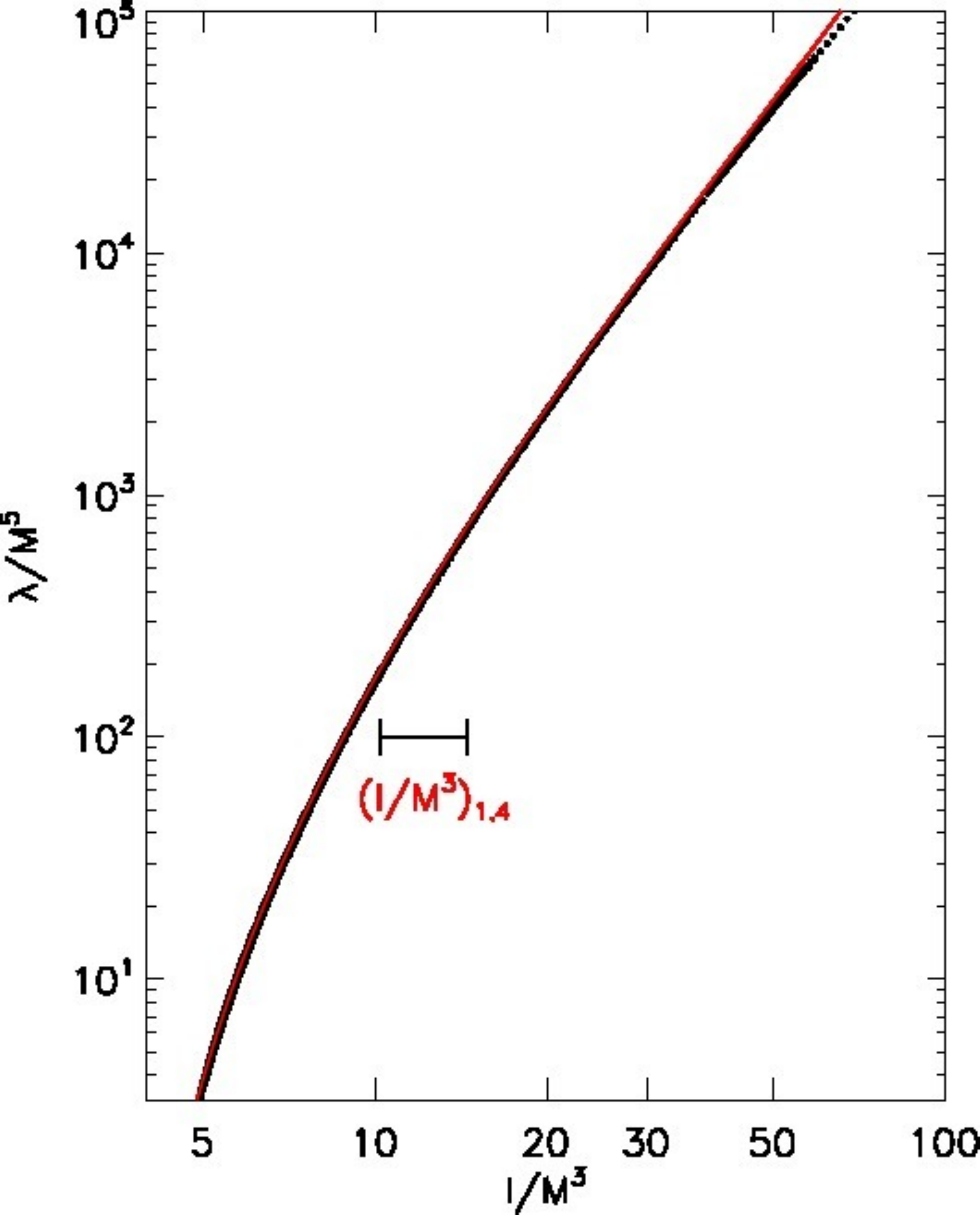}
  \caption{The I-Love relation between $\bar I=I/M^3$ and
    $\bar\lambda=\lambda/M^5$. Black circles show results for the
    piecewise polytropic EOSs, the red line is Eq. (\ref{eq:lam}).}
  \label{fig:lami}
\end{figure}

\begin{figure}
  \includegraphics[width=8.5cm]{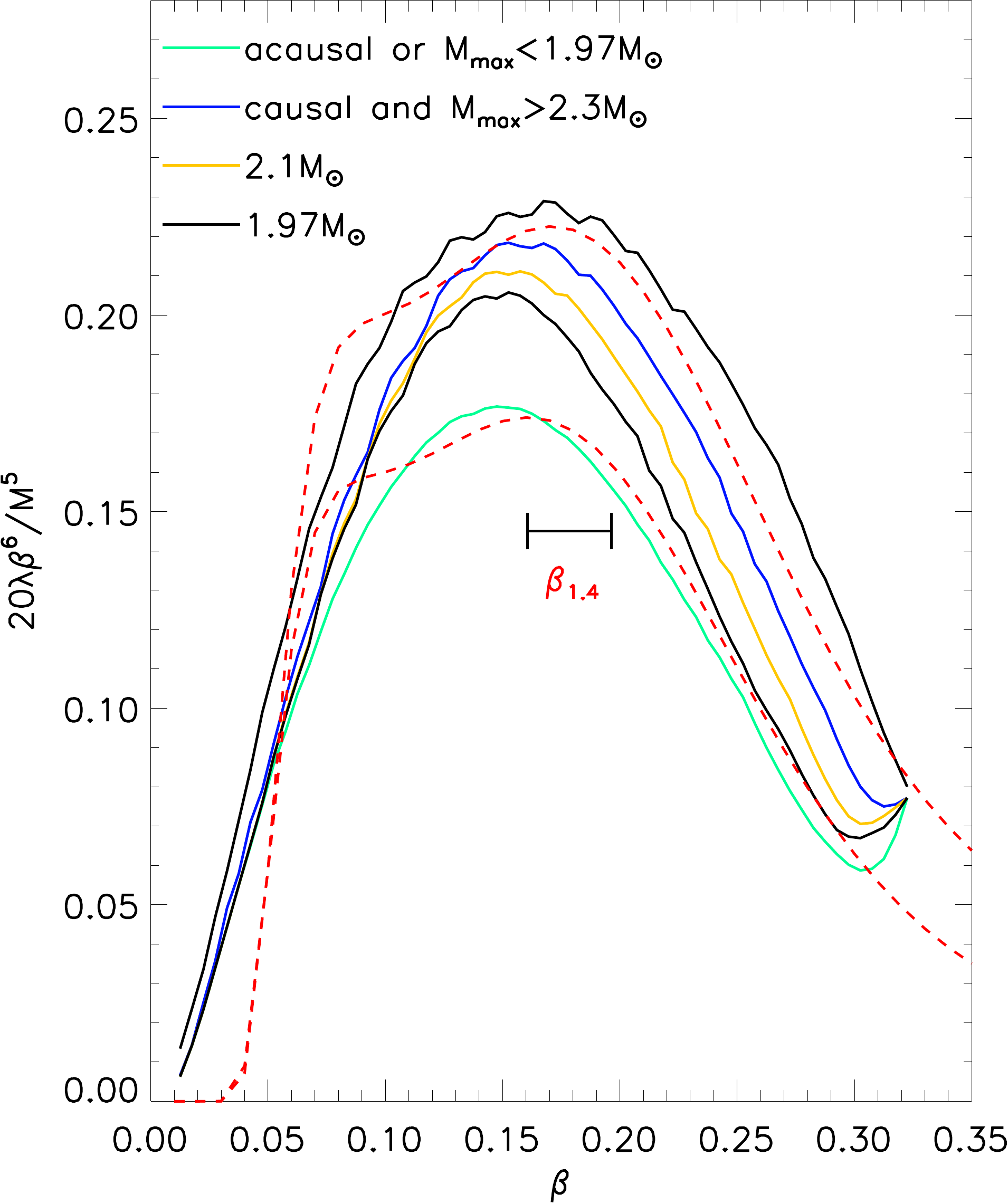}
  \caption{Similar to Fig. \ref{fig:be} but for the dimensionless
    tidal Love number as a function of compactness. The dashed lines
    show the combined approximations from Eqs. (\ref{eq:lam}) and
    (\ref{eq:i}) .}
  \label{fig:lam}
\end{figure}

\begin{figure}
  \includegraphics[width=8.5cm]{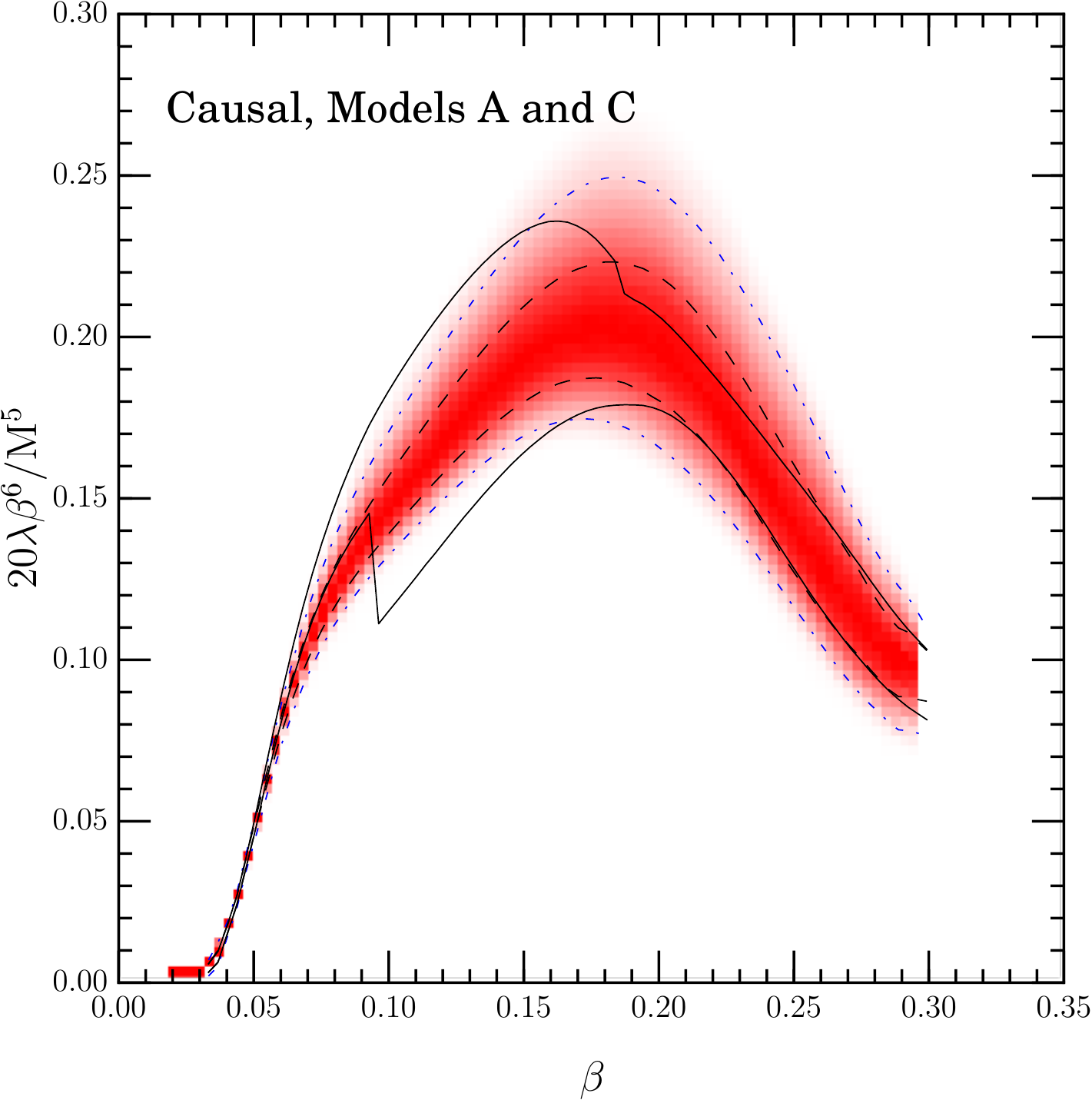}
  \caption{The quantity $20 \lambda \beta^6/M^5$ as a function of
    compactness for models A and C. The 68\% and 95\% contour lines
    are indicated with the same notation as
    Fig.~\ref{fig:R14P1p85both} above. Note that the correlation is
    only weakly dependent on the prior distribution. The jump in the
    contours for Model C at $\beta=0.1$ is a result of selecting a new
    peak in a slightly bimodal distribution.}
  \label{fig:lb6_beta_both}
\end{figure}

Ref.~\cite{Yagi13} found an e\-ven more re\-mark\-a\-ble
EOS-in\-de\-pen\-dent re\-la\-tion relating the moment of inertia, the
tidal Love number and the quadrupole polarizability, which is now
known as the I-Love-Q relation. The correlation between the
dimensionless moment of inertia, $\bar I=I/M^3$, and the dimensionless
tidal Love number, $\bar\lambda=\lambda/M^5$, is shown in Fig.
\ref{fig:lami} for the piecewise polytropic EOSs. The relation for
$\bar I(\bar\lambda)$~\cite{Yagi13} and its inverse, $\bar\lambda(\bar
I)$, are
\begin{align}
  \ln\bar I & \simeq1.417+0.0817\ln\bar\lambda+
  0.0149(\ln\bar\lambda)^2\\
  &+0.000287(\ln\bar\lambda)^3
  -0.0000364(\ln\bar\lambda)^4,\\
  \ln\bar\lambda&\simeq-30.5395
  +38.3931\ln\bar I-16.3071(\ln\bar I)^2\\
  &+3.36972(\ln\bar I)^3-0.26105(\ln\bar I)^4.
\label{eq:lam}
\end{align}
The deviation of piecewise polytropic EOSs from these analytic
approximations is negligible, as seen in Fig. \ref{fig:lami}.
Combining Eq. (\ref{eq:lam}) with the approximation Eq. (\ref{eq:i})
yields an explicit approximation for $\bar\lambda(\beta)$, as shown in
Fig. \ref{fig:lam}. The dashed lines show the combined approximations
from Eqs. (\ref{eq:lam}) and (\ref{eq:i}). It is apparent that
uncertainties in a revised correlation would be substantially reduced
if observational constraints for $M_{\mathrm{max}}$ were taken into
account.

\begin{figure}
  \includegraphics[width=8.5cm]{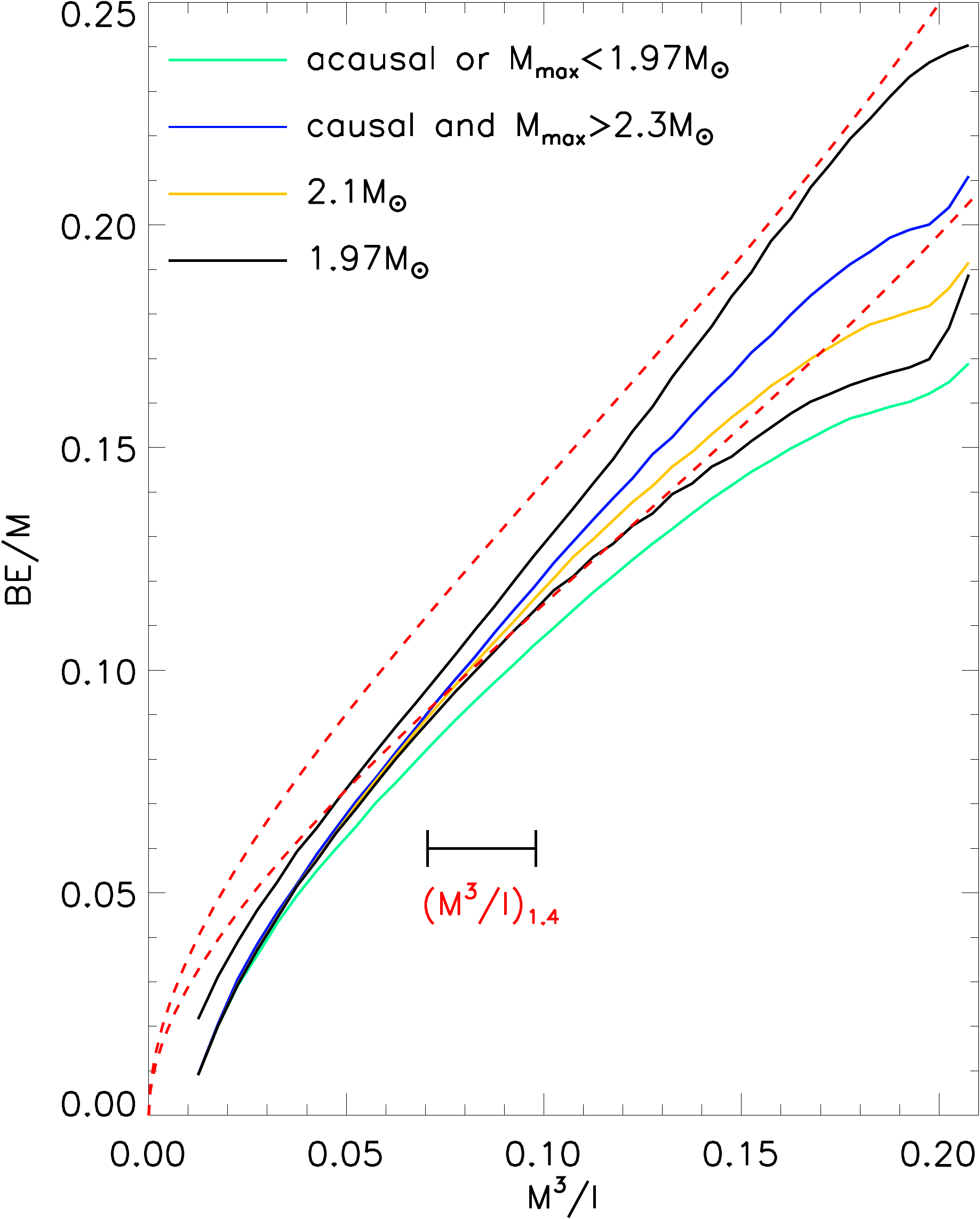}
  \caption{Similar to Fig. \ref{fig:be} but for the dimensionless
    binding energy BE/$M$ as a function of $1/\bar I$. The dashed
    lines represent the combination of approximations for $I(\beta)$
    and BE$(\beta)$, Eqs. (\ref{eq:i}) and (\ref{eq:bex}). }
  \label{fig:bei}
\end{figure}

It is interesting to examine the EOS dependence of the moment of
inertia relative to the binding energy (Fig. \ref{fig:bei}). The two
quantities are not as highly correlated as are $\bar I$ and
$\bar\lambda$, but the correlation is still significant. The
approximation shown in Fig. \ref{fig:bei}, using
$M_{\mathrm{max}}>1.97~\Msun$, is
\begin{align}
  {\rm BE}/M& = 0.0075+\left(1.96^{+0.05}_{-0.05}\right) \bar{I}^{-1}
  -12.80 \bar{I}^{-2} \\
&+72.00\bar{I}^{-3}-\left(160^{+20}_{-20}\right)\bar{I}^{-4}
\label{eq:bei}
\end{align}
The improvement resulting from inclusion of maximum mass constraints,
compared to the correlation ${\rm BE}(\beta)/M$ inferred from Eqs.
(\ref{eq:i}) and (\ref{eq:bex}), is seen to be substantial, especially
for compactnesses characteristic of $1.4~\Msun$ stars.

\section{Discussion}

In the future, astronomical observations of neutron stars may be
sufficiently plentiful and precise that one will be able to map out
the mass-radius curve and the EOS using simpler methods like a
$\chi^2$ analysis. The currently available observational data does not
permit this. The problem of determining the mass-radius curve is
underconstrained, in part because the data is not yet precise, and in
part because there are several remaining systematic uncertainties (or
several model assumptions).

A similar difficulty exists on the side of experimental nuclear
physics. Even though isospin-asymmetric matter near the saturation
density is well-known, the nature of neutron-rich matter near the
nuclear saturation density is still subject to uncertainties such as
the nature of the three-neutron force and the highest density to which
we can trust chiral effective theories (see recent progress on this
front in Ref.~\cite{Furnstahl15}. Experimental observables which probe
high-density matter, such as in\-ter\-me\-di\-ate-en\-er\-gy heavy-ion
collisions (e.g. Ref.~\cite{Tsang09co}), are also subject to strong
systematics. As we have noted above, the lowest density at which a
phase transition to exotic matter is possible is a critical quantity
necessary for understanding neutron stars. This lowest density limit
for exotic matter is not well-known.

In the context of Bayesian inference, we approached these issues by
varying our prior assumptions. To the extent that we are able to
express the various systematic uncertainties in terms of prior
probability distributions, Bayesian inference provides us a way to
{\em quantify} the uncertainty in the mass-radius curve and the EOS of
dense matter. This method also allows us to critically examine the
extent to which correlations between observables exist.

We found, in particular, that the possible presence of phase
transitions at low-density has an important impact on the lower-limit
for the radius of low-mass neutron stars. We showed that some
universal relations (or correlations) are not strongly sensitive to
the prior assumptions, such as the relation between the radius of a
$1.4~\Msun$ neutron star, the pressure at $\approx 2 n_s$, and the
correlation between \Mmax and the pressure at $\approx 4 n_s$. We also
showed that the high degree of correlation between the moment of
inertia and the tidal Love number (the I-Love correlation) is robust
with respect to prior assumptions concerning the equation of state.
Finally, we presented a new universal relation connecting the neutron
star binding energy to its moment of inertia.

\begin{acknowledgement}
  AWS was supported by the U.S. DOE Office of Nuclear Physics. JML was
  supported by the U.S. DOE grant DE-AC02-87ER40317 and the DOE
  Topical Collaboration ``Neutrinos and Nucleosynthesis in Hot and
  Dense Matter''. EFB was supported by the NSF under grant No.'s
  PHY-1430152 (JINA Center for the Evolution of the Elements) and AST
  11-09176. This project used computational resources from the
  University of Tennessee and Oak Ridge National Laboratory's Joint
  Institute for Computational Sciences.
\end{acknowledgement}

\bibliographystyle{prsty}
\bibliography{epja}

\end{document}